\documentclass[12pt]{article}
\usepackage{graphicx}
\usepackage{epsfig}
\usepackage{epsf}
\usepackage{amssymb}
\usepackage{rotate}

\newcommand{ \slashchar }[1]{\setbox0=\hbox{$#1$}   
   \dimen0=\wd0                                     
   \setbox1=\hbox{/} \dimen1=\wd1                   
   \ifdim\dimen0>\dimen1                            
      \rlap{\hbox to \dimen0{\hfil/\hfil}}          
      #1                                            
   \else                                            
      \rlap{\hbox to \dimen1{\hfil$#1$\hfil}}       
      /                                             
   \fi}                                             %
\def\ptmiss{\slashchar{p}_{T}}

\usepackage{cite}

\usepackage[dvips]{color}
\usepackage[normalem]{ulem}

\definecolor{darkgreen}{cmyk}{1,0,1,0.4}

\newcommand{\bea}{\begin{eqnarray}}
\newcommand{\eea}{\end{eqnarray}}

\def\bmu {B_{\mu}^{(1,0)}}
\def\bh {B_H^{(1,0)}}

\def\ltap{\raisebox{-.4ex}{\rlap{$\sim$}} \raisebox{.4ex}{$<$}} 
 
\def\beq{\begin{equation}} 
\def\eeq{\end{equation}} 
\def\barr{\begin{array}}
\def\earr{\end{array}}

\def\gev{\, {\rm GeV}} 
\def\tev{\, {\rm TeV}}

\begin{document}

\vskip 15pt

\begin{center}
{\Large \bf Exploring two Universal Extra Dimensions at the CERN LHC} \\
\vspace*{1cm}
\renewcommand{\thefootnote}{\fnsymbol{footnote}}
{ {\sf Debajyoti Choudhury$^1$}, {\sf Anindya Datta${}^{2}$},
{\sf Dilip Kumar Ghosh$^{3}$} and {\sf Kirtiman Ghosh${}^{4}$}
} \\
\vspace{10pt}
{\small ${}^{1)}$ 
{\em Department of Physics and Astrophysics, University of Delhi,
Delhi 110 007, India} \\
   ${}^{2)}$ {\em Department of Physics, University of Calcutta,
92 A.P.C. Road, Kolkata 700 009, India}}  \\
   ${}^{3)}$ {\em Department of Theoretical Physics, 
         Indian Association for the Cultivation of Science,\\
          2A \& 2B Raja S.C. Mullick Road, Kolkata 700 032, India}\\
   ${}^{4)}$ {\em Harish-Chandra Research Institute,
Chhatnag Road, Jhunsi, Allahabad  211 019, India}
\normalsize

\abstract 
We discuss the signatures,
at the LHC, of the (1,0)-th Kaluza-Klein (KK)
gluon and quarks in the framework of the two universal extra
dimensional (2UED) model. Once produced,
these particles typically suffer a cascade of decays terminating
in the Dark Matter candidate apart from Standard Model particles.
 In this article, we are interested in focus on
 a particular signature of 2UED  wherein the final state
comprises of at least four jets in association with a hard photon
and missing transverse energy. Several kinematic cuts are devised to
enhance the signal to background ratio. Finally, as a road map to
parameter determination at the LHC, we point out an interesting
correlation between the peak position of the $M_{eff}$ distributions
with the compactification radius $R$ and the cut-off scale $M_s$.\\

\end{center}

\section{Introduction}

Apart from finding out the so far elusive Higgs boson, the other main
aim for the Large Hadron Collider (LHC) experiments is to explore any
new dynamics operative at the TeV regime. As of now, the Standard
Model (SM) remains very successful in explaining almost all of the
experimental data related to elementary particle physics.  Some
discrepancies (mostly in flavour physics) do remain though, and while
the statistical significance of each may not be a overriding cause of
concern, together, they point to the tantalizing prospect of some new
physics being just around the corner.  Moreover, certain other
inadequacies of the SM remain, e.g. the hierarchy problem, the lack of
a Dark Matter candidate, an understanding of neutrino masses, a lack
of sufficient baryogenesis etc. Such issues have led to a plethora of
new physics models being proposed.  In this endeavour, models defined
in more than three spatial dimensions need special attention.

Originally, extra dimensions had been proposed as a way to achieve
unification of gravity and electrodynamics.  Although, the initial
constructions were beset with problems and soon rendered irrelevant,
in later years, extra dimensions found a natural place in string
theory. However, traditionally such extra dimensions were sought to be
compactified with a radius too small to be of any phenomenological
consequence.  On the other hand, in recent years, (large) extra
dimensions have been invoked to solve the hierarchy problem of the SM.
In the models proposed by Arkani-Hamed, Dimopoulos and Dvali
\cite{ArkaniHamed:1998rs} and Randall and Sundrum
\cite{Randall:1999ee}, for example, the SM is confined to a sub-space
of the $1 + (3+\delta)$ dimensional manifold, while gravity can
propagate into all of the $4+ \delta $ dimensions.  On the contrary,
in another class of models, some or all the SM fields can propagate
into space beyond the usual $1+3$ dimensions.  While the name
Universal Extra Dimensional (UED) model \cite{Appelquist:2000nn}
strictly applies only to the case wherein each of the fields can
percolate to all of the dimensions, the usage is often expanded to
include scenarios wherein at least some of the SM fields propagate in
more than four dimensions.  Apart from a very rich phenomenology
(which can be probed by the LHC), these models offer gauge coupling
unification at a relatively low scale of energy \cite{Dienes:1998vg,
  Dienes:1998vh,Bhattacharyya:2006ym}, and naturally contain a weakly
interacting massive stable particle, which can be a suitable candidate
for cold dark
matter\cite{Servant:2002aq,Kong:2005hn,Kakizaki:2006dz,Belanger:2010yx}.
While the simplest UED scenario would have only one extra dimension
({\em minimal UED} or mUED model), a particularly interesting variant
is the eponymous {\em two Universal Extra Dimension (2UED)}
model \cite{Burdman:2005sr,Dobrescu:2004zi}.  By definition, all the
SM particles propagate in the entire $1+3+2$--dimensional space-time.
Apart from providing a cold dark matter candidate
\cite{Dobrescu:2007ec}, this model naturally suppresses proton decay
rate to below the current constraints\cite {Appelquist:2001mj} as well
as predicts the number of fermion generations \cite{Dobrescu:2001ae},
the last features rendering it superior to the minimal UED model.
Although other variants of the 2UED model have been proposed
since~\cite{Dohi:2010vc}, we shall restrict ourselves to the simplest
one.

For the sake of simplicity, let us assume that the two extra
space--like dimensions have the same size.  Furthermore, let us assume
that the gauge structure as well as the particle content is the
same\footnote{Note that such a construction would be untenable in the
  mUED as parity is undefined in five dimensions.} as in the
SM.  Consequently, after compactification (discussed in detail in a
following section), the physical spectra will contain,
apart from the SM particles,
respective double towers of Kaluza-Klein (KK) excitations,
with each excitation being specified by two integers, 
$(i,j)$, called the KK-numbers. In addition, the 5th and 
6th components of the gauge bosons
will appear in the low-energy theory as scalars 
transforming under the respective adjoint representations. 
Phenomenology of
these spinless adjoints have been investigated in detail in Refs.
\cite{Burdman:2006gy,Dobrescu:2007xf,Freitas:2007rh,Ghosh:2008ix,Ghosh:2008dp,
Ghosh:2008ji}.

In this article, we will discuss, instead, search strategies for the
strongly interacting $(1,0)$-mode particles, as their production is
favoured at the LHC.  Once produced, they will eventually decay to
(1,0) mode EW gauge bosons/quarks 
along with SM quarks.  The former decay, in turn, producing more stable
SM particles like leptons, photons and quarks. Finally, the decay
cascade terminates at the production of $B_H ^{(1,0)}$, the first
scalar excitation of the $U(1)$ gauge boson.  This, being the lightest
$(1,0)$ mode particle, cannot decay further due to the conservation of
KK-parity. So, in general, production of $(1,0)$ mode strongly
interacting particles at the LHC will be characterised by the presence
of number of leptons/jets/photons in association with transverse
missing energy due to the weakly interacting $B_H ^{(1,0)}$. In this
article, we will be interested in a particular signature of this type,
namely, $n$-jets (with $n \ge 4$) $+$ single photon $+$ missing
transverse energy.  The somewhat complementary signal 
comprising of multi-leptons plus
missing energy at the LHC as well as at the ILC has been investigated
in Ref. \cite{Ghosh:2008ji}.

The plan of the present article is as  following. In section 2, 
we will discuss the 2UED model in general and particularly the
SM in $1 + 5$ dimensions. Section 3 will be devoted to the
phenomenology of the 
$(1,0)$-level of this model. In the next section, we
will present our main result, i.e., the search strategies at the LHC
operating with a center of mass energy of 7 and 14 TeV. Finally, we 
conclude in section 5.

\section{Two Universal Extra Dimensions}

In this section, we will briefly introduce the 2UED model, 
wherein all the SM fields can
propagate universally in the $(1+3+2)$--dimensional space-time.  With
$x^{\mu}$ ($\mu=0,1,2,3$) denoting the Minkowski space, the
compactification of the two extra dimensions can be described as
follows:

\begin{itemize}
\item 
The flat extra dimensional space (before orbifolding) is a square with
sides $L$, viz. $0\le x^4,x^5\le L$
\cite{Burdman:2005sr,Dobrescu:2004zi}\footnote{{In accordance 
with  Refs.\cite{Burdman:2005sr,Dobrescu:2004zi}, we}
have chosen the size of the
extra dimensions  {to be} the same. 
However, the most general case would
imply two different sizes for these two directions. In absence of any
obvious symmetry that relates these two length scales, we are thus
considering only a specific choice.}.
Identifying the opposite sides of the square would make the
compactified manifold a torus. However, toroidal compactification
leads to 4D fermions that are vector-like with respect to any gauge
symmetry. The alternative is to identify two pairs of adjacent sides
of the square \cite{Burdman:2005sr,Dobrescu:2004zi}, namely,
\begin{equation}
 (y,0)~\equiv~(0,y),~~~(y,L)~\equiv~(L,y),~~~\forall~y\in~[0,L] \ .
\label{b_cond}
\end{equation}
This is equivalent to folding the square along a diagonal and glueing
the boundaries.  The above mechanism automatically leaves at most a
single 4D fermion of a given chirality as the zero mode of any chiral
6D fermion \cite{Dobrescu:2004zi}.

\item 
Clearly, the identification of Eq.~(\ref{b_cond}) is valid only 
if, for any pair of identified points,
the Lagrangian assumes identical value for any field
configuration, viz.
\begin{eqnarray}
{\cal L}\vert_{x^\mu,y,0}\;=\;{\cal L}\vert_{x^\mu,0,y};~~
{\cal L}\vert_{x^\mu,y,L}\,=\,{\cal L}\vert_{x^\mu,L,y} \ .
 \nonumber
\end{eqnarray}
This requirement fixes the boundary conditions for 6D scalar fields
and Weyl fermions.  While the gauge kinetic term allows for a two-fold
ambiguity, the requirement that the boundary conditions for 6D scalar
or fermionic fields be compatible with the gauge symmetry also fixes
the boundary conditions for 6D gauge fields.

\item Any 6D field (fermion/gauge
 or scalar) $\Phi(x^\mu,x^4,x^5)$ can be decomposed as

\begin{equation}
 \Phi(x^\mu,x^4,x^5)~=~\frac{1}{L}\sum_{j,k}f^{(j,k)}_n(x^4,x^5)
\Phi^{(j,k)}(x^\mu),
\label{kk_dcomp}
\end{equation}
where,
\begin{equation}
 f^{(j,k)}_n(x^4,x^5)~=~\frac{1}{1+\delta_{j,0}\, \delta_{k,0}}
\left[e^{-in\pi/2}\, \cos\left(\frac{jx^4+kx^5}{R}
    + \frac{n\pi}{2}\right)+ 
   \cos\left(\frac{kx^4-jx^5}{R}+\frac{n\pi}{2}\right)\right] ,
\label{kk_func}
\end{equation}
with the compactification radius $R \equiv L / \pi$.  The 4D fields
$\Phi^{(j,k)}(x^\mu)$ are the $(j,k)$-th KK-modes of the 6D field
$\Phi(x^{\alpha})$ and $n$ is an integer whose value is restricted to
$0,1,2$ or $3$ by the boundary conditions.

\item 
The functions $f^{(j,k)}_n(x^4,x^5)$ should form a complete set on the
compactified manifold, and, thus, must satisfy
\begin{equation}
\frac{1}{L^2}\sum_{j,k}\left[f_{n}^{(j,k)}(x^4,x^5)
  \right]^*f_{n}^{(j,k)}(x^{\prime4},x^{\prime 5}) ~=~\delta(x^{\prime
4} - x^{4}) \delta(x^{\prime 5} - x^{5}). 
\label{comp}
\end{equation}
It is clear from the form of $f_{n}^{(j,k)}$ that the functions
$f_{n}^{(1,0)}$ and $f_{n}^{(0,1)}$ are not independent
($f_{n}^{(0,1)}=(-1)^n f_{n}^{(1,0)}$). Therefore, it is sufficient to
take the set ($j > 0,~k \ge 0$) along with $j=k=0$ to form a complete
set of functions on the chiral square.  It is also obvious from the
form of $f^{(j,k)}_n(x^4,x^5)$ that only $n=0$ allows zero mode
($j=k=0$) fields in the 4D effective theory.
The zero mode fields and the interactions among zero modes can
be identified with the SM.

\item In 6D, the Clifford algebra is generated by six anticommuting
matrices, $\Gamma^{\alpha},~ \alpha=0,1,..,5$, 
with the minimum
dimensionality of the matrices being $8\times8$.
Akin to 4D, 
the spinor
representation of the $SO(1,5)$ Lorentz symmetry is reducible and
contains two irreducible Weyl representation characterized by different
eigenvalues of the 6D chirality operator $\bar \Gamma \equiv
\Gamma^0\Gamma^1\Gamma^2\Gamma^3\Gamma^4\Gamma^5$. 
The chirality projection operators are defined as
$P_{\pm}=(1\pm\bar\Gamma)/2$, where $+$ and $-$ label the 6D
chiralities defined by the eigenvalues of $\bar \Gamma$,  {viz.}
\begin{equation}
\Psi_{\pm}(x^{\alpha})= P_{\pm}\Psi(x^\alpha) \ , \qquad
 \bar \Gamma \Psi_{\pm}
(x^\alpha) = \pm \Psi_{\pm}(x^{\alpha}).
\label{chirality}
\end{equation}
The chiral fermions in 6D have four components. Each 6D chiral fermion
contains both the chiralities of $SO(1,3)$.
\end{itemize}

\subsection{Introduction to the SM in 6D (6DSM)}

Now we move on to the Standard Model in 6-dimensions. In 6D, the fields
and boundary conditions are chosen such that, upon compactification
and orbifolding,
the zero modes of the resulting effective theory  reproduce the
SM. The requirements of anomaly cancellation and fermion mass generation
 force the weak-doublet fermions to have opposite {\em 6D chiralities}
 with respect to the weak-singlet fermions. So the quarks of one
 generation are given by $Q_+~\equiv~(U_+,D_+),~U_-,~D_-$. Since
 observed quarks and leptons have definite 4D chirality, an immediate
 constraint is imposed on the boundary conditions of doublet and
 singlet fermions. 
 The 6D doublet quarks and leptons decompose into a  {double} tower
 of heavy vector-like 4D fermion doublets with left-handed zero
 mode doublets.  Similarly, each 6D singlet quark and lepton decomposes
 into towers of heavy 4D vector-like singlet fermions along with
 zero mode right-handed singlets. The
 zero mode fields are
 identified with the SM fermions. 
For example, SM doublet and
 singlets of 1st generation quarks are given by
 $(u_{L},d_{L})~\equiv~Q_{+L}^{(0,0)}(x^{\mu})$, 
$u_{R}~\equiv~U_{-R}^{(0,0)}(x^{\mu})$
 and $d_{R}~\equiv~D_{-R}^{(0,0)}(x^{\mu})$.

A given gauge field $A_\alpha \; (\alpha = 0, 1 \dots 5)$, on
compactification, decomposes into a tower of 4D spin-1 field, with
each of the two additional degrees of freedom {\em viz.} $A_4$ and
$A_5$ giving rise to an additional tower of 4D spin-$0$ fields. One
combination of the latter is eaten by the former to yield heavy spin-1
fields. The other combination remains in the physical spectrum as a
tower of {\em spinless adjoints}.  For example, the
6D gluon fields, $G_{\alpha}^{a}$ decompose into a tower of 4D spin-1
fields, $G_{\mu}^{a(j,k)}$, and a tower of spin-0 fields,
$G_{H}^{a(j,k)}$, with the former including a zero mode to be identified with the SM gluon.  Similarly, the 6D $SU(2)$
gauge fields have KK-modes $W_{\mu}^{(j,k)\pm}$, $W_{H}^{(j,k)\pm}$,
$W_{\mu}^{(j,k)3}$ and $W_{H}^{(j,k)3}$, while the hypercharge gauge
field has KK-modes $B_{\mu}^{(j,k)}$ and $B_{H}^{(j,k)}$. The zero
modes of $W_{\mu}^{(j,k)\pm}$ towers are identified with the SM
$W^{\pm}_{\mu}$ bosons. The mixing of $W_{\mu}^{(0,0)3}$ and
$B_{\mu}^{(0,0)}$ gives the photon and the $Z$-boson.  {For} non-zero
modes this mixing is negligible, though.

The tree-level masses for $(j,k)$-th KK-mode particles are given by
$\sqrt{M_{j,k}^2~+~m_{0}^{2}}$, where $M^2_{j,k} \equiv (j^2+k^2)/R^2$
and $m_0$ is the mass of the corresponding zero mode particle. As a
result, for sufficiently large $R^{-1}$ (as demanded by
phenomenological consistency), the tree-level masses are approximately
degenerate within a given non-trivial KK level. This degeneracy is
lifted by radiative effects \cite{Ponton:2005kx}.  The fermions
receive mass corrections from the gauge interactions (with gauge
bosons and spinless adjoints) as well as Yukawa interactions, with
each contributing a positive mass shift.

The gauge fields and spinless 
adjoints receive mass corrections from the self-interactions and gauge
 interactions. 
While the fermion loops (via gauge interactions) lead to negative mass shifts,
self-interactions give positive shifts. The latter are, of course,
nonexistent for the $B_{\mu}^{(j,k)}$ and for the corresponding
scalars $B_{H}^{(j,k)}$.  Explicit computations show that the lightest
KK particle is the spinless adjoint $B_{H}^{(1,0)}$.  As a result, the
2UED model gives rise to a scalar dark matter \cite{Dobrescu:2007ec}.
Consistency of the relic abundance of the $B_{H}^{(1,0)}$ with 
$\Omega_{DM}$ as inferred from WMAP observations \cite{wmap} leads to 
a very constrained allowed region in the $R$--$m_{\rm higgs}$ plane, 
confined to $R^{-1} \ltap 500 \gev$. However, note that the analysis 
of Ref.\cite{Dobrescu:2007ec} was performed only at the leading order
and higher order corrections would, typically, be expected to increase
the annihilation cross-sections and, thereby, push up the allowed 
range in $R^{-1}$. Much more importantly, the inclusion of the higher 
KK modes in the annihilation (as well as co-annihilation) of the DM
candidate significantly raises the $R^{-1}$ range, as has recently 
been demonstrated~\cite{Belanger:2010yx} for the mUED case. In the
current context, this effect is expected to be even stronger. Consequently,
we shall not impose the WMAP constraint on the maximum allowed value
for $R^{-1}$.

\section{Phenomenology of $(1,0)$-mode sector of 6DSM}
In the preceding section, we identified the standard model doublet and
singlet quarks with the $(0,0)$ modes of the 6D fields $Q_+$, $U_-$
and $D_-$ respectively. Similar would be the case for the leptonic
fields. The $(1,0)$--mode fermionic sector thus constitutes of
$Q_+^{(1,0)}$, $U_-^{(1,0)}$, $D_-^{(1,0)}$, $L_+^{(1,0)}$ and
$E_-^{(1,0)}$. As for the corresponding bosonic sector, we have, apart
from the Higgs (scalar) and gauge boson (vector) excitations, another
set of three scalars transforming under the adjoint representation of
the respective gauge groups. The last mentioned, for which there is no
analog in the mUED case, would play a key role.

\begin{figure}[t]
\begin{center}
\epsfig{file=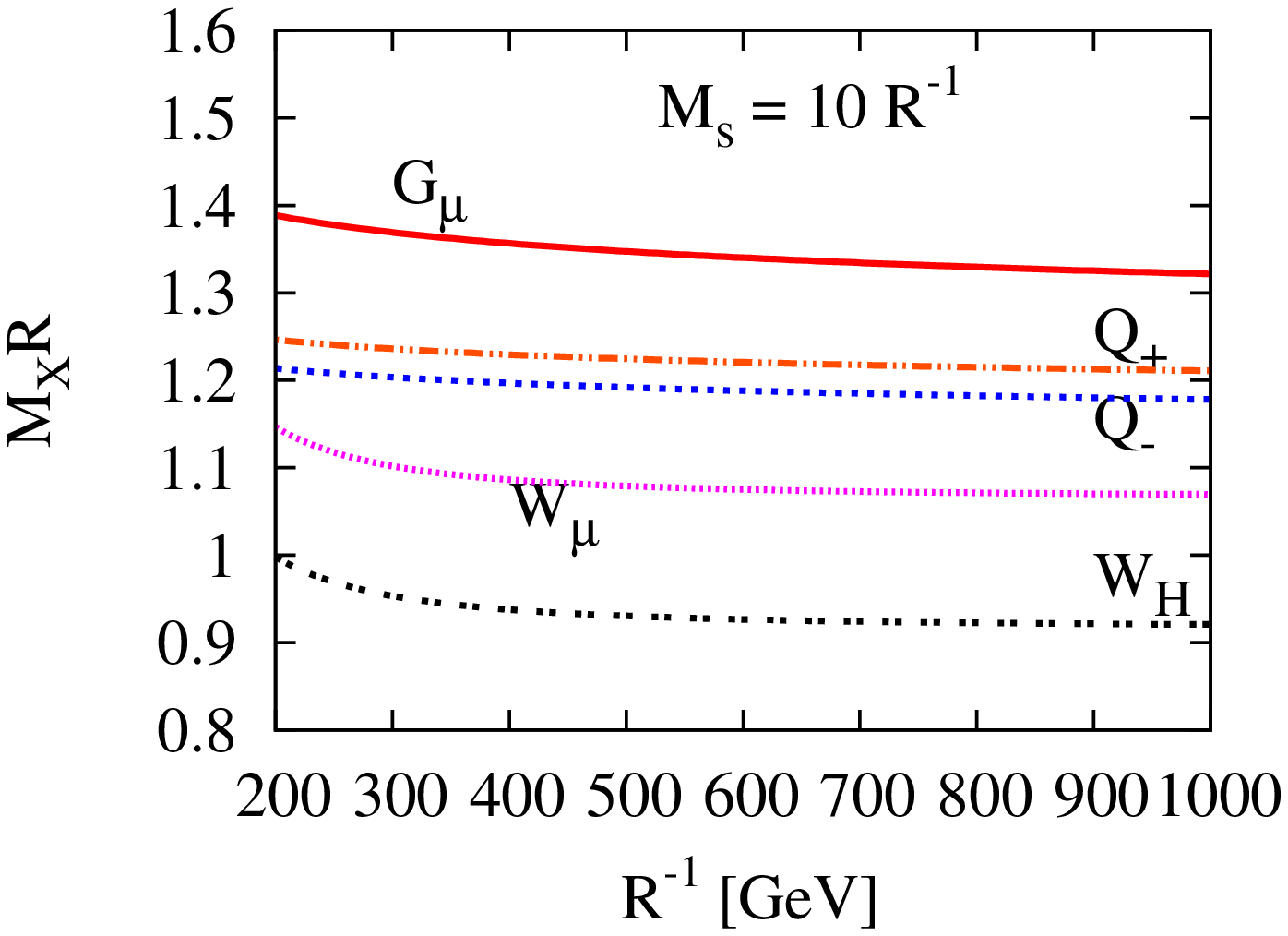,width=6cm,height=8cm,angle=270}
\epsfig{file=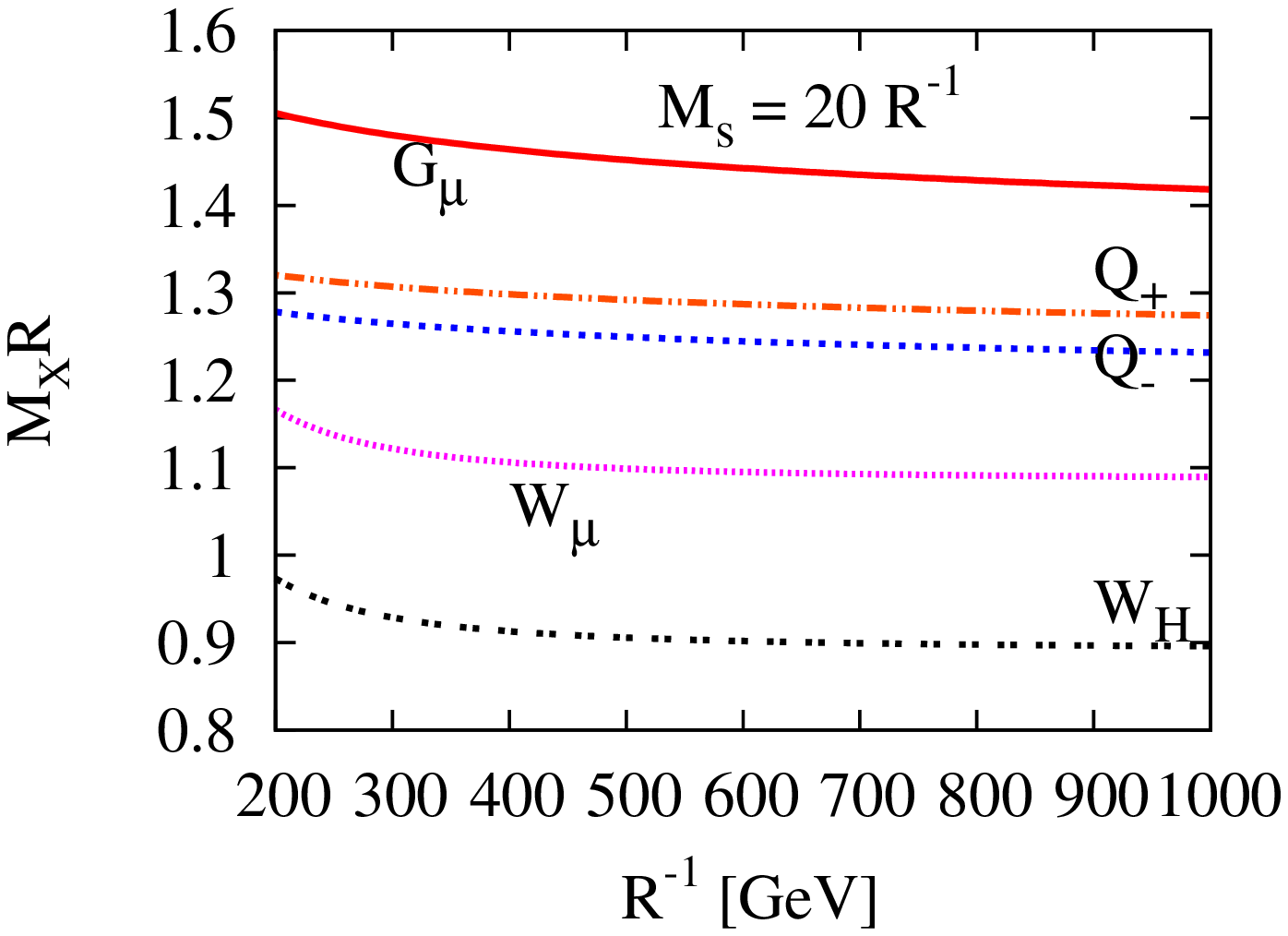,width=6cm,height=8cm,angle=270}
\end{center}
\caption{Variation of $M_XR$ (where $X$ corresponds to either
  $G_\mu^{(1,0)},~Q_+^{(1,0)}$, $Q_-^{(1,0)}$, $W_\mu^{(1,0)}$ or
  $W_H^{(1,0)}$) as a function of $R^{-1}$ for two different values of
  {$M_s \,R$. Here, $Q_\pm^{(1,0)}$ does not include the top's
    partners.}}
\label{fig:mass}
\end{figure} 
While the tree-level masses, in the absence of electroweak 
symmetry breaking, would be $R^{-1}$ for each, the inclusion of radiative 
corrections does change them\cite{Ponton:2005kx}. 
With the change being dependent on the cutoff scale $M_s$ 
(note that a ultraviolet completion needs 
to be defined for all such theories), we present  
the corrections for $M_s=10~(20)R^{-1}$. To be specific, 
\bea  
M_{L_{+}}&\simeq&{1.04}(1.06)~{R^{-1}}, \hskip 50pt
M_{E_{-}}~\simeq~1.0 (1.0){R^{-1}}, 
\nonumber\\ 
M_{B_{\mu}}&\simeq&{0.97} (0.96)~{R^{-1}}, \hskip 50pt
M_{B_{H}}~\simeq~{0.86} (0.82)~{R^{-1}}, 
\nonumber\\ 
&~&~~~~~~~~~~~~ M_{G_H}~\simeq~{1.0}(1.0)~{R^{-1}},
\label{mass}
\eea 
with the numerical factors being almost independent of $R^{-1}$.   
For the  {other} colored states, an additional mild dependence accrues
from the scale dependence of the QCD coupling constant. 
For
the $SU(2)$ gauge bosons and spinless adjoints, the 
 {$R^{-1}$}
dependence arises from the non-zero mass of the SM $W^{\pm}$ and $Z$-boson. 
In Fig.~\ref{fig:mass}, we present these masses as a
function of $R^{-1}$ with 
 {$\alpha_s = \alpha_s (M_X)$.} It should be noted that a smaller $M_sR$  
leads to a more degenerate spectrum, and, consequently, in the 
collider context, a more difficult situation to explore.

\subsection{Decay of $(1,0)$-mode particles}
Decays of  {the} $(1,0)$-mode particles have been 
investigated in detail in Ref.~\cite{Dobrescu:2007xf}.  Conservation
of KK-parity allows $(1,0)$-mode particles to decay only into a
$(1,0)$-mode particle and one or more SM particles if kinematically
allowed. It is clear from Eq. (\ref{mass}) that $B_{H}^{(1,0)}$ is the
lightest KK particle (LKP) in this theory. It is important to notice
that, unlike 
the case of the mUED, the LKP is now 
a scalar.
Since the $\bh$ is
a stable particle and weakly interacting, it passes through the
detector without being detected. Decays of all the $(1,0)$-mode
particles thus result in one or more SM particles plus missing
energy/momentum signature. We briefly discuss below the decays of the
different $(1,0)$-mode particles. 
Unless specified, we shall, henceforth, limit ourselves to 
$300 \gev \le R^{-1} < 1 \tev$ and $M_s = 10 \, R^{-1}$. Furthermore,
we shall neglect the production of 
$T_{+,-}^{(1,0)}$, the partners of the top quarks.
 
\begin{itemize}
\item \underline{\bf $(1,0)$-mode gluons ($G_\mu^{(1,0)}$):} 
The heaviest of the $(1,0)$-mode particles, the
$G_\mu^{(1,0)}$ has tree-level gauge couplings
  with a SM quark and the corresponding $(1,0)$-mode quark.
The decay of $G_\mu^{(1,0)}$ into $Q_{-}^{(1,0)}$ is slightly
favoured by phase space. The branching fractions of
$G_\mu^{(1,0)}$ into a quark plus $Q_{+}^{(1,0) i}$, $U_{-}^{(1,0)
  i}$, or $D_{-}^{(1,0) i}$, summed over the index $i$ which labels
the three generations, are  36.7\%, 24.6\% and 38.7\%,
respectively. It is important to 
note that, for $1/R \le 1.3$ TeV,
the decays of the $(1,0)$ vector gluon into $t_L T_{+}^{(1,0) }$ or
$t_R T_{-}^{(1,0) }$ are  {kinematically} forbidden.

\item \underline{\bf $(1,0)$-mode quarks ($Q_+^{(1,0)}$ and
  $Q_-^{(1,0)}$)}: These are heavier than the $(1,0)$--mode
  electroweak gauge bosons and {all of the} spinless adjoints, and
  can, thus, decay into either, accompanied by a quark.
  Understandably, the decay to the $SU(3)$-spinless adjoint (driven by
  QCD) is the dominant one for both doublet and singlet $(1,0)$-mode
  quarks. The $SU(2)_W$ doublet (1,0) quarks can also decay into a SM
  quark, and an $SU(2)_W$ gauge boson or spinless adjoint. Both
  doublet and singlet $(1,0)$-mode quarks may also decay into
  $(1,0)$-mode hypercharge bosons or spinless adjoints. The branching
  fractions of $Q_{+}^{(1,0)}$ into
  $qG_{H}^{(1,0)},~qW_{H}^{(1,0)3}(W_{H}^{(1,0)+})$ and
  $qW_{\mu}^{(1,0)3}(W_{\mu}^{(1,0)+})$ are given by 63.2\%,
  5.6\%(11.2\%) and 6.4\%(12.8\%) respectively.  The decay of the doublet
  quarks into $q B_{H}^{(1,0)}(B_\mu^{(1,0)})$ is suppressed by the
  hypercharge.  The branching fractions of $Q_{-}^{(1,0)}$ into $q_R
  G_H^{(1,0)}$ and $q_R B_{\mu}^{(1,0)}(B_H^{(1,0)})$ {are} given by
  82.1\% and 11.5\%(6.4\%) respectively for the up type quarks
  ($U_{-}^{(1,0)}$) and 94.8\% and 3.3\%(1.9\%) respectively for the
  down type quarks ($D_{-}^{(1,0)}$).

\item \underline{\bf $SU(3)_C$ spinless adjoint ($G_H^{(1,0)}$)}: This
  suffers a tree-level 3-body decay into a $q \bar q$ pair plus one of
  the electroweak $(1,0)$-mode gauge boson or spinless adjoint.  The
  dominant mode is $G^{(1,0) }_H \to B^{(1,0)}_H q\overline{q}$, with
  a total branching fraction of $96.5\%$. The other decay modes 
   {(into electroweak gauge bosons and spinless adjoints)} are
  suppressed by the small mass splitting between  {them and the} 
$G^{(1,0) }_H$.

\item \underline{\bf $SU(2)$ gauge bosons {($W^{(1,0)\pm}_\mu~{\rm
      and}~W^{(1,0)\, 3}_\mu$)}:} Since these are heavier than the
  $(1,0)$-mode leptons, they decay dominantly into
the doublet fields. For 
  example, $W^{(1,0)3}_\mu$ can decay
  into one of the six ($l_i L_{+i}^{(1,0)}$ and $\nu_i
  \nu_{+i}^{(1,0)},~i=e,\mu,\tau$) channels with equal
  probability. Similarly, $W^{(1,0)\pm}_\mu$ decays into one of the
  six possible
  modes ($l_i \nu_{+i}^{(1,0)}$ and $\nu_i
  L_{+i}^{(1,0)},~i=e,\mu,\tau$) with 
  a branching fraction of $1/6$ into
  each.

\begin{figure}
\begin{center}
\epsfig{file=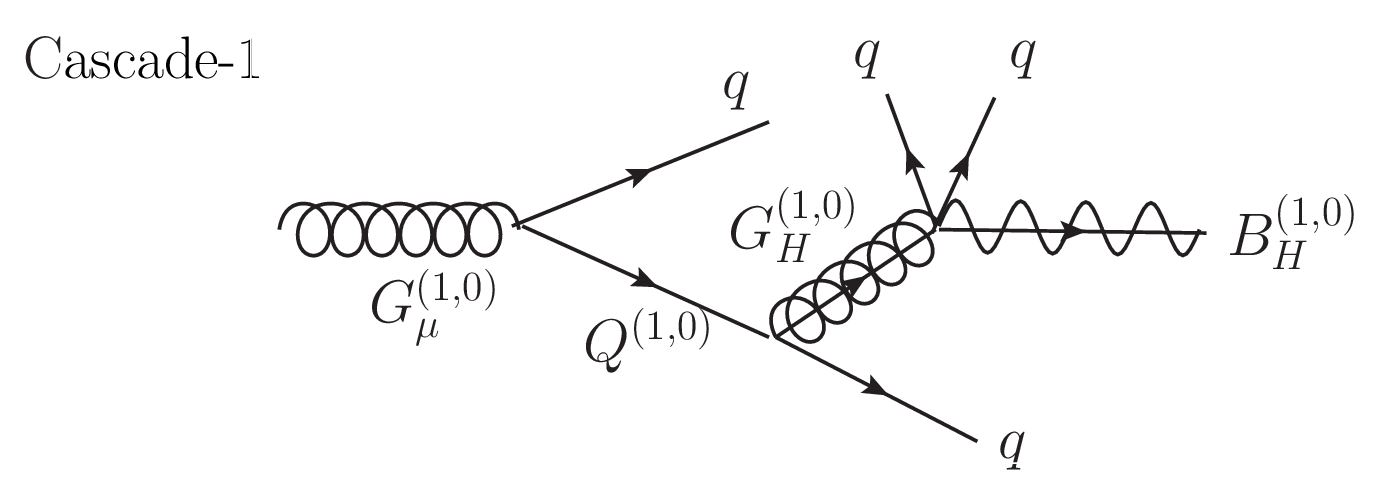,width=8cm,height=4cm,angle=0}
\epsfig{file=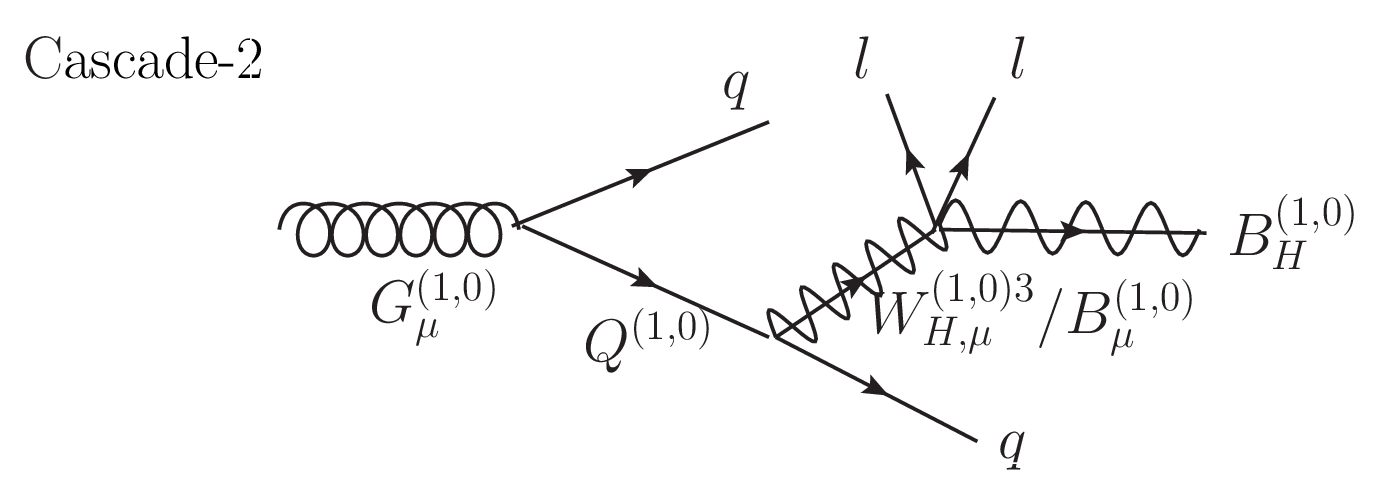,width=8cm,height=4cm,angle=0}\\
\vskip 10pt \epsfig{file=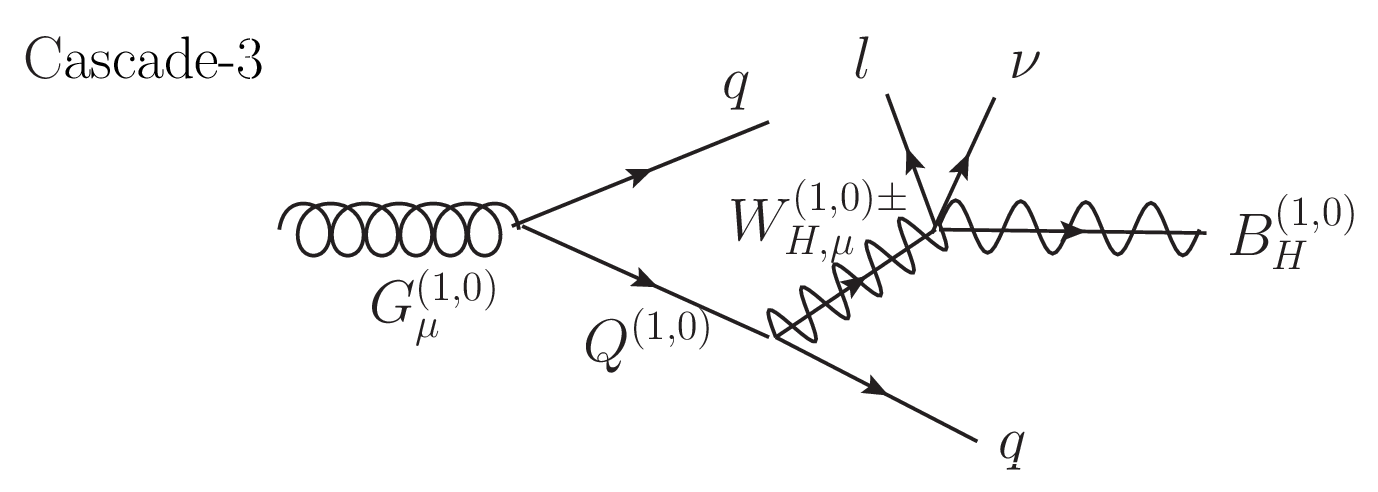,width=8cm,height=4cm,angle=0}
\epsfig{file=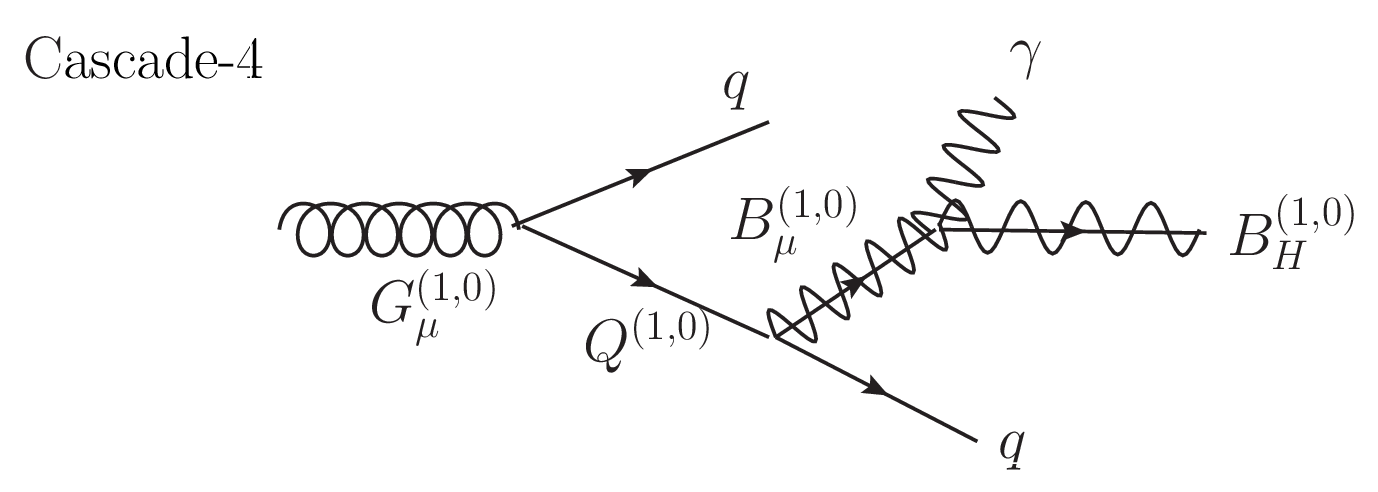,width=8cm,height=4cm,angle=0}
\end{center}
\caption{Decay cascades of $G_\mu^{(1,0)}$.} 
\label{fig:cascead}
\end{figure}

\item \underline{\bf $(1,0)$-mode leptons ($L_{+}^{(1,0)}$ and
  $E_{-}^{(1,0)}$):} Being heavier than $B_\mu^{(1,0)}$, $B_H^{(1,0)}$ and
  $SU(2)$ spinless adjoints, they can decay into the latter and the
  corresponding lepton.

\item \underline{\bf $U(1)$ gauge boson ($\bmu$):} 
It has a tree level 3-body decay into a pair of SM fermions and the $\bh$. 
Being driven by the hypercharge, the decay into the right-handed 
leptons 
 {dominates} that into the doublet fermions. 
Since the corresponding $(1,0)$ fermion
appears in the propagator, the decays into quarks 
(e.g., $\bmu~\to ~u_R \bar u_R \bh$) suffer 
a further suppression on account of the largeness of the propagator mass.
On the other hand, the one-loop decay $\bmu \to \gamma \bh$ amplitude 
receives contribution from each of the fermion species as well as
the Higgs. While this 
amplitude would have vanished in the limit of 
all the $(1,0)$ fermionic fields being degenerate, 
the split described above is sufficient to render it 
quite substantial.
Indeed, the dominant modes are 
$\bmu$ in $l \bar l \bh$ (where $l$ includes $e,~\mu$
and $\tau$) and $\gamma \bh$ with the respective branching fractions, 
for $R^{-1}=500$ GeV, being 63.5\% and 36.2\%.

\item \underline{\bf $SU(2)$ spinless adjoints ($W^{(1,0)\pm}_H$ and
  $W^{(1,0)\, 3}_H$):}
These can decay only to the $B_{H}$ and SM particles
\cite{Ghosh:2008ix}.  The $W^{(1,0)\, 3}_H$ decays to a pair of SM
  leptons and $\bh$ with equal branching ratio to charged leptons and
  neutrinos. Branching fraction to quark antiquark pairs is again
  negligible due to hypercharge and heavy $(1,0)$--mode quark
  propagator.  The $W^{(1,0)\pm}_H$, on the other hand, decay with
  almost 100\% branching ratio to $ l \bar\nu_{l} \bh$ ($l$ includes
  all 3 SM lepton generations). Branching fractions of $SU(2)$
  spinless adjoints are essentially independent of $R^{-1}$.

\end{itemize}

We are now equipped enough to discuss the decay cascade of the
strongly interacting $(1,0)$-mode particles. Let us first begin with
the $G_\mu^{(1,0)}$, which decays into a SM quark and the
corresponding $(1,0)$-mode quarks.  The latter decays into the
spinless adjoints or electroweak gauge bosons in association with a SM
quark.  The decay of $Q_-^{(1,0)}$ into $G_{H}^{(1,0)}$ results in
four SM quarks plus $B_{H}^{(1,0)}$ in the final state,
whereas the decay
into $SU(2)$ gauge bosons or spinless adjoints gives
rise to two SM quarks and two SM leptons in association with a
$B_{H}^{(1,0)}$.  However, there is one more interesting decay mode
available, 
 namely $Q_-^{(1,0)}\to q \bmu$,
which
 gives rise to two SM quark $+$ one photon
$+$ a $B_{H}^{(1,0)}$ at the end of 
a single
$G_\mu^{(1,0)}$ decay cascade. The
decay cascades of $G_\mu^{(1,0)}$ are schematically shown in
Fig. \ref{fig:cascead}.

\subsection{Production cross sections}
Owing to KK-parity, the $(1,0)$ mode particles 
can only be pair-produced. We shall restrict ourselves 
only to the production of strongly interacting
$(1,0)$-mode particles as the cross-sections for 
the color-singlet states are suppressed by more than 
an order of magnitude. 

All the $(1,0)$-mode gauge bosons and spinless adjoints have tree
level couplings with an $(1,0)$-mode fermion and a SM fermion
{arising} from the compactification of the 6D kinetic term for
fermions.  Similarly, the compactification of the kinetic term for the
6D gluon field gives rise to both $G_{\mu/H} G_{\mu/H} g$ and
$G_{\mu/H}G_{\mu/H}gg$ couplings. Thus, $GG$ production will proceed
from both the $gg$ initial state (by virtue of the aforementioned
couplings) as well as from a $q \bar q$ initial state (through a
$s$-channel SM gluon). Owing to the larger mass of the $G$, this
particular mode never dominates though. What does, for a significant
part of the parameter space, is $Q G$ production which proceeds from
the initial $q g$ state through a combination of three Feynman
diagrams (a $s$-channel $q$, a $t$-channel $Q$ and a $u$-channel
$G$). Also of particular interest is $q_i q_j \to Q_i Q_j$ production
that proceeds through $t/u$-channel $G$ exchange. Given that the LHC
is a $pp$ machine, this would be expected to dominate for large
$R^{-1}$.

While the electroweak diagrams would also contribute to the 
last-mentioned (as well as to $Q_i \, \bar Q_j$ production), 
these amplitudes are suppressed by a relative factor of 
$(\alpha_{EW}/\alpha_s)$. And given that they do not bring in 
any new topologies, their 
total contribution is rather subdominant.

\begin{figure}[!h]
\begin{center}
\epsfig{file=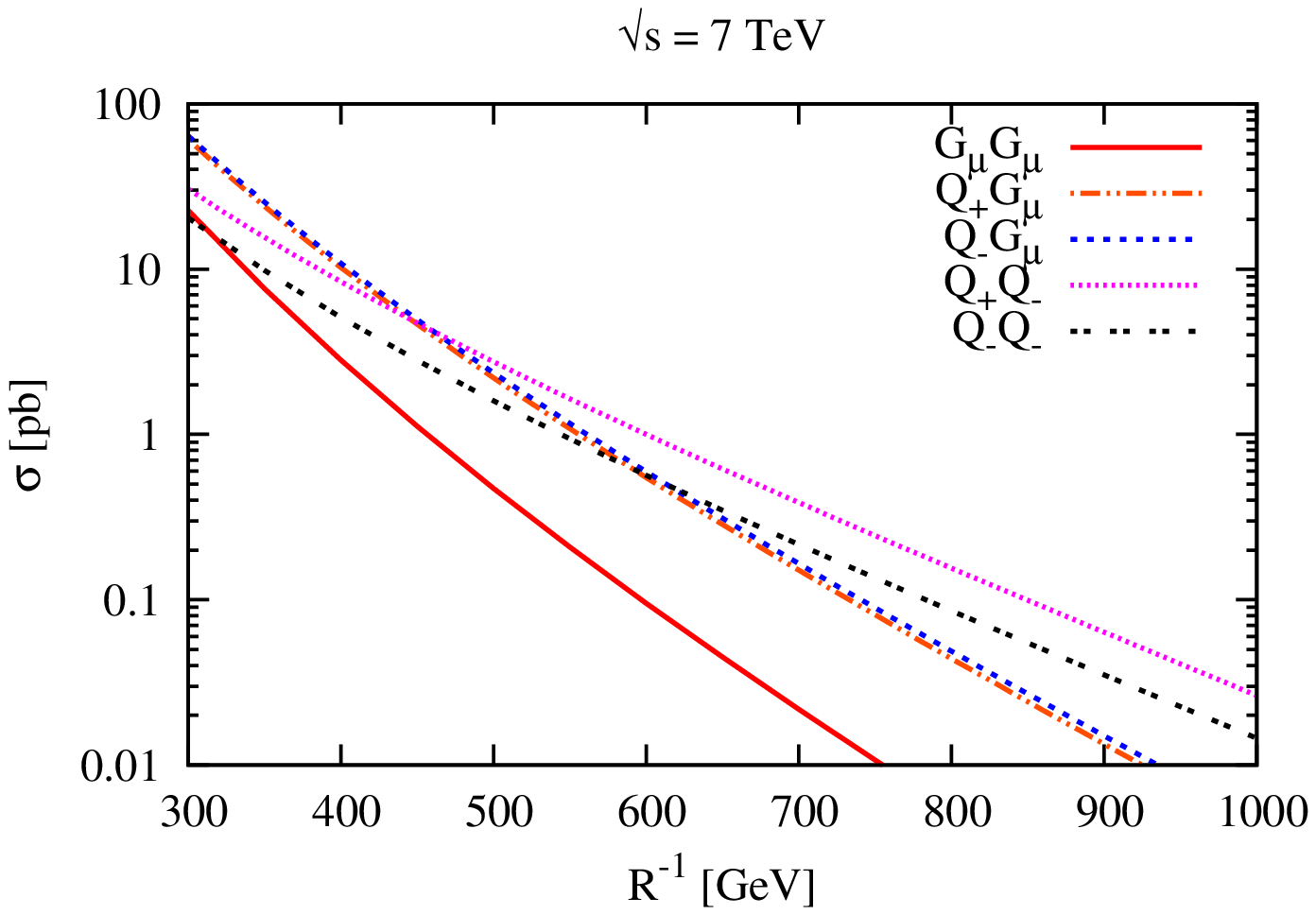,width=7cm,height=8cm,angle=-90}
\epsfig{file=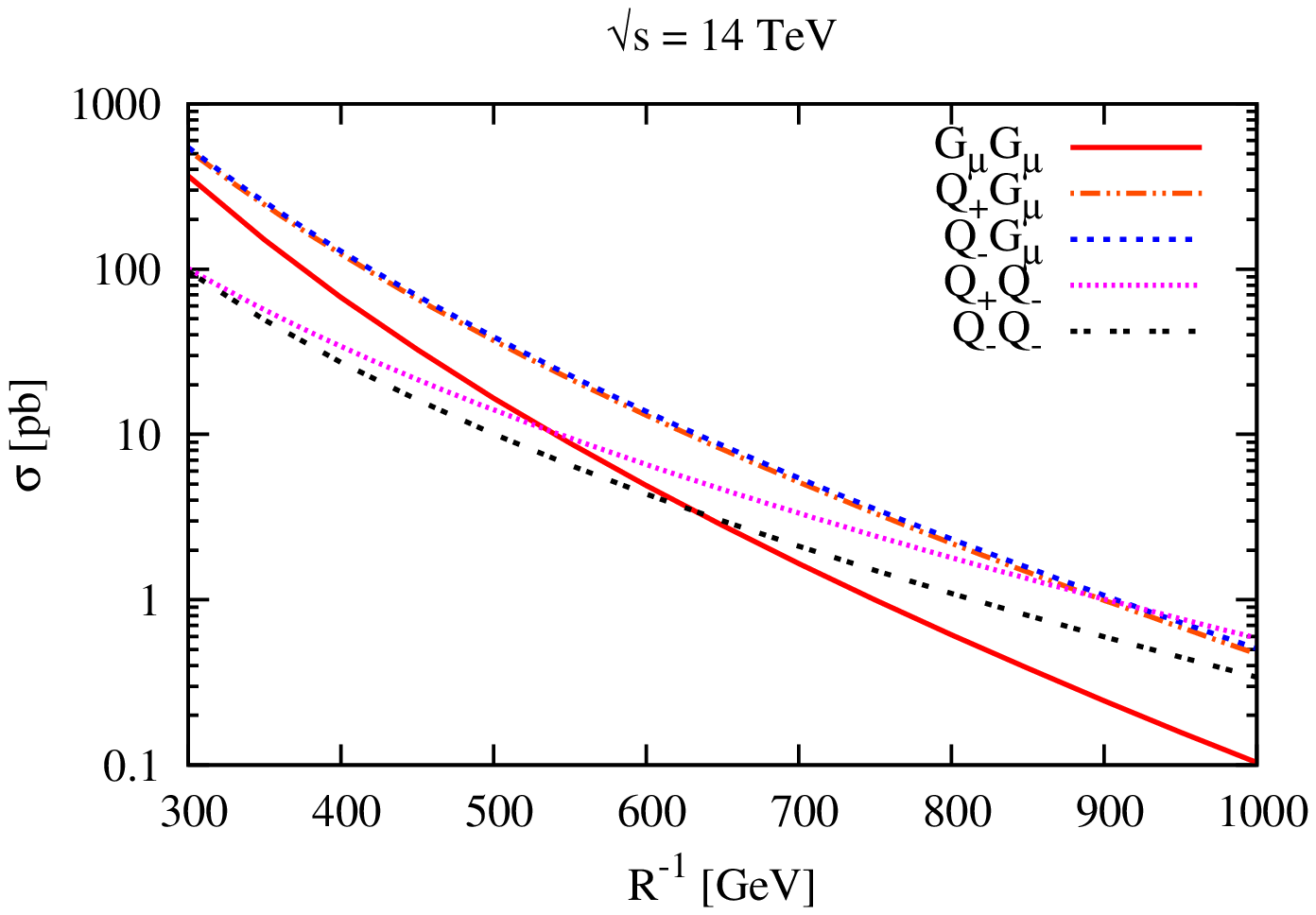,width=7cm,height=8cm,angle=-90}
\end{center}
\vspace*{-4ex}
\caption{Total pair production cross-sections of $(1,0)$-mode
  particles at the LHC operating with $\sqrt s=$ 7 TeV (left), and 14 TeV
  (right).}
\label{fig:cross}
\end{figure}
Finally, the processes of interest are 
\begin{equation}
pp \Longrightarrow
G_\mu + G_{\mu}, \quad
G_\mu + Q_{i}, \quad
Q_{i} + Q_{j},  \quad
Q_{i} + \bar Q_{j} \ , 
\end{equation}
where the indices $i, j$ run over both flavour and 
the 6D chirality. In Fig.~\ref{fig:cross}, we show 
the total cross sections for some of these modes 
at the LHC, obtained using the CTEQ6L parton 
distribution functions 
\cite{Pumplin:2002vw,Stump:2003yu} 
with the factorization scale fixed at $ Q^2 = \hat s/4$ (for $300 {\rm
~GeV} \leq R^{-1} \leq 1000 \rm ~GeV$).  While there exists a dependence on these two choices, we
are not, by any means, overestimating the signal size.  In the absence
of any computation of the
higher order corrections (expected only to enhance these numbers),
we limit ourselves to only tree-level calculations.

\section{Signature of 6DSM at the LHC}
In general, pair production of coloured $(1,0)$ -mode particles---
depending upon the decay chain---can give rise to
$n$-jet+$m$-lepton+$\ell$-photon+$\ptmiss$ signatures.
For example, pair production of the
$SU(3)_C$ gauge boson $G_\mu $, followed by its hadronic decay, 
leads to 8 partons accompanied by missing-$\ptmiss$, 
with each parton potentially leading to one jet (or more).
However, the multi-jet+$\ptmiss$
  signature is expected to be overwhelmed by the pure QCD
  background\footnote{Although Ref. \cite{Murayama:2011hj} has
    claimed, in a different context, that such a final state can be
    used, it requires very sophisticated handling and the robustness
    of the stratagem developed therein is yet to be vindicated.}, and
  we shall desist from using it any further.

  Multi-jet+multi-lepton+$\ptmiss$
results when one $G_\mu$ follows the decay ``Cascade-1'' of
Fig. \ref{fig:cascead} and other follows either of
``Cascade-2'' and 
``Cascade-3''. It has been shown in Ref.~\cite{Bhattacherjee:2010vm}
that this
signature can serve
as a potential discovery channel for the mUED model.  In principle, one
could perform a similar kind of analysis in the context of {\em 6DSM}
as well.
However, the presence of $G_H^{(1,0)}$ in the particle spectrum 
significantly reduces the leptonic branching fractions for the 
excited quarks and gluons when compared to the mUED.

Instead, we consider the final state
\beq
n \mbox{--jets} + \gamma+ p_T\!\!\!\!\!\!/ 
 \qquad (n \geq 4)
   \label{signal}
\eeq
where the hard photon will be used 
as an additional trigger.
The main motivations for choosing this particular signal 
topology are  two fold. First of all, 
this signature serves to distinguish between the 
2UED and mUED models.
The particle spectrum of the mUED model does not contain the 
spinless adjoints.
Hence, the first KK-excitation of the $U(1)$ gauge boson is the LKP
therein, and its further decay is not possible.
The second
reason is of course the rate of the SM background, which is 
reduced substantially with the emission of an 
isolated hard photon in the multi-jet events. 
In the 2UED model, such a
final state topology arises 
when one $G_\mu$ decays 
hadronically and the other one follows the 
``Cascade-4'' decay chain.

\begin{figure}[t]
\begin{center}
\epsfig{file=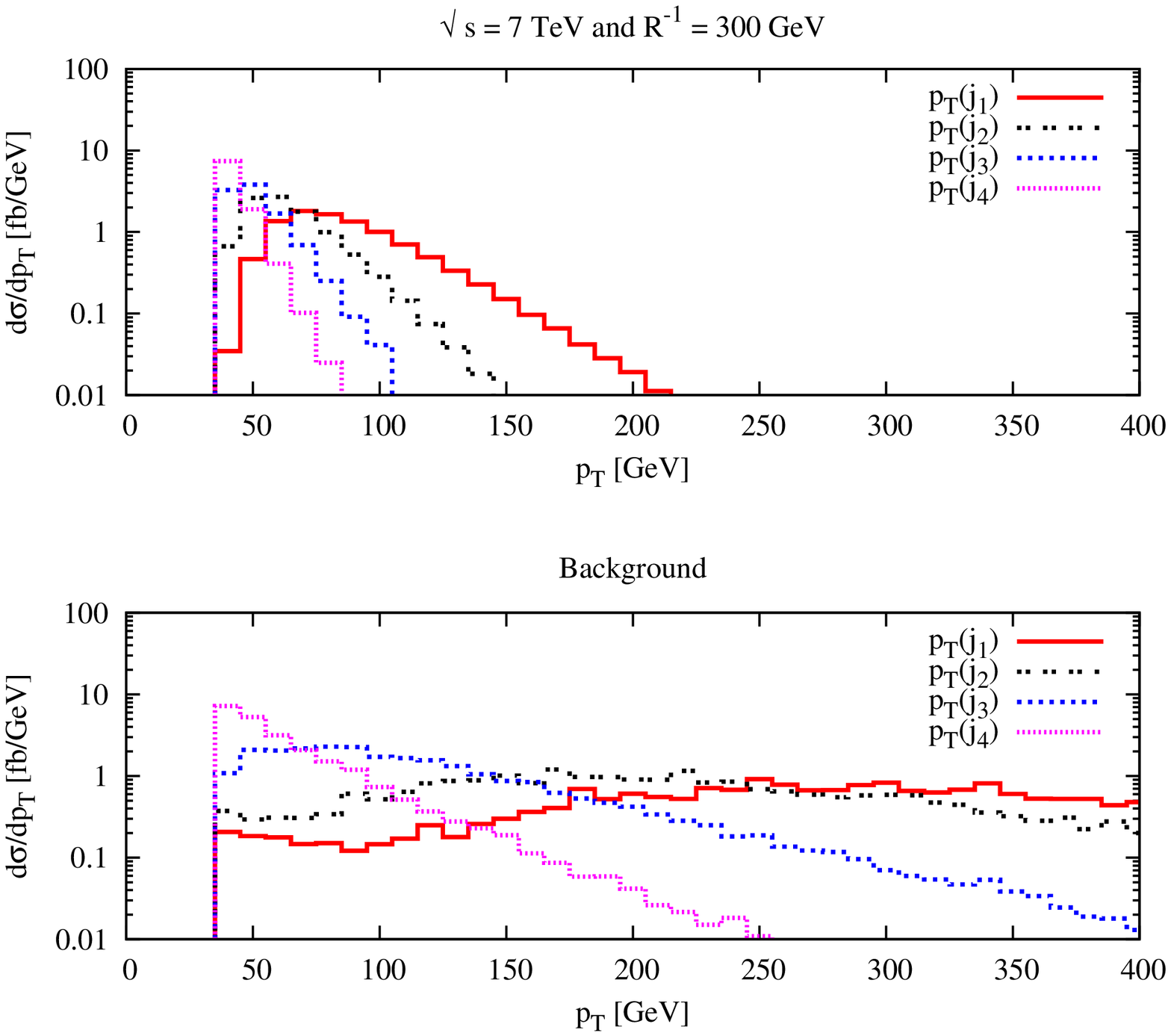,width=12cm,height=8cm,angle=-90}
\epsfig{file=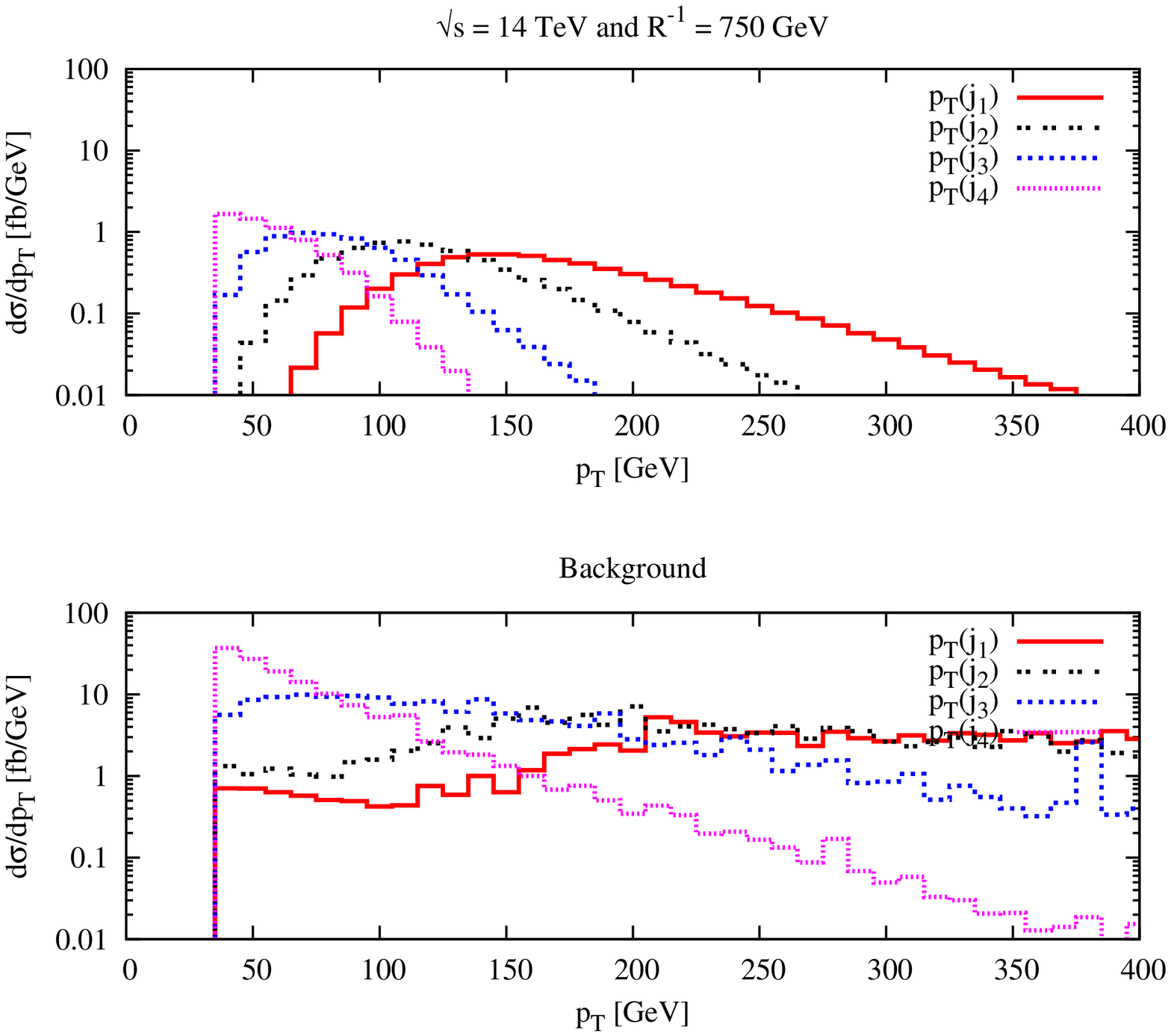,width=12cm,height=8cm,angle=-90}
\end{center}
\caption{Transverse momentum distribution of jets 
(after ordering them according to their $p_T$ 
hardness) after the acceptance cuts for both signal 
(top panel) and the SM background 
(bottom panel) at the LHC with $\sqrt s=$ 7 TeV (left), 
and 14 TeV (right).} 
\label{fig:jetpt}
\end{figure}
\begin{figure}[t]
\begin{center}
\epsfig{file=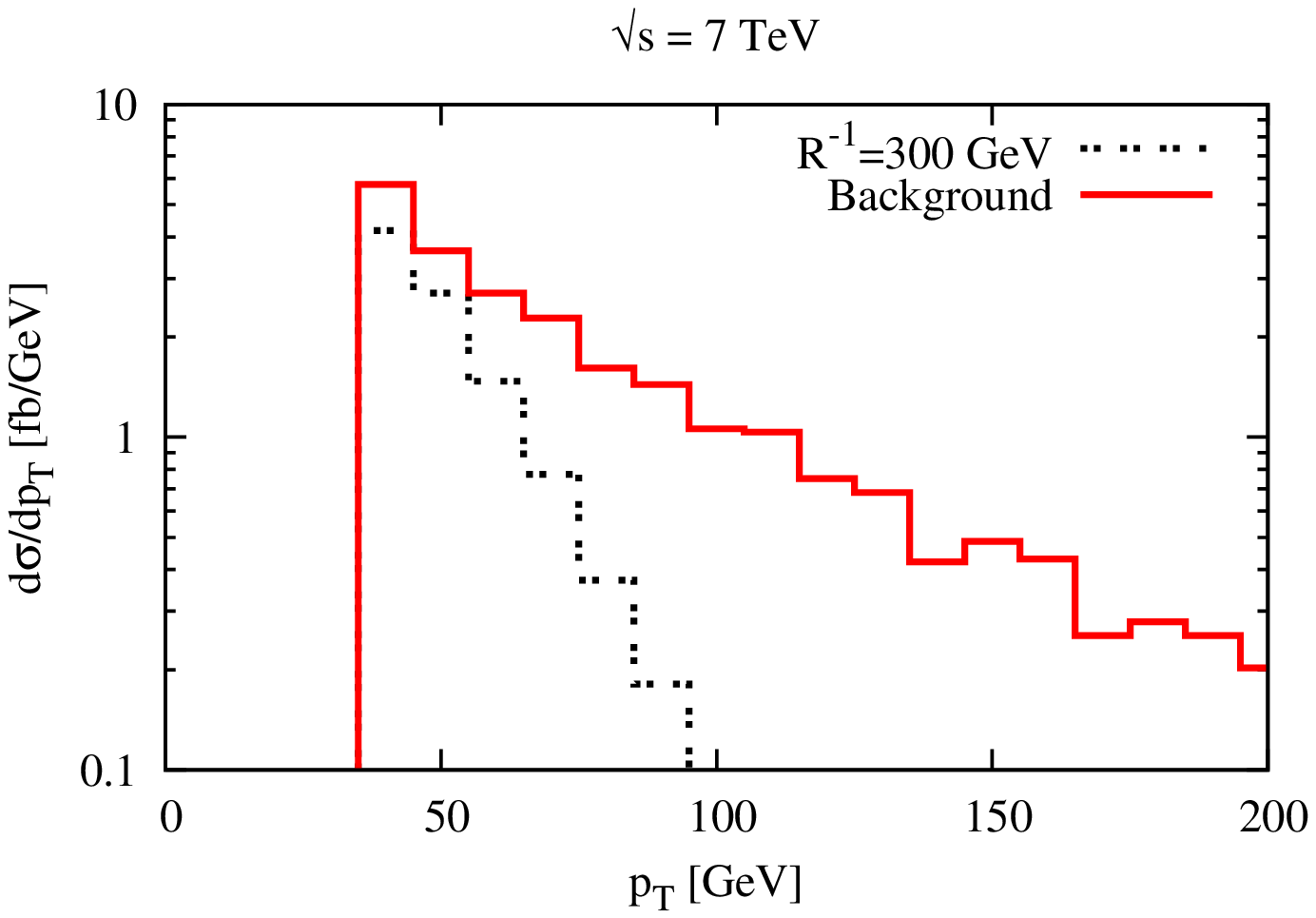,width=8cm,height=8cm,angle=-90}
\epsfig{file=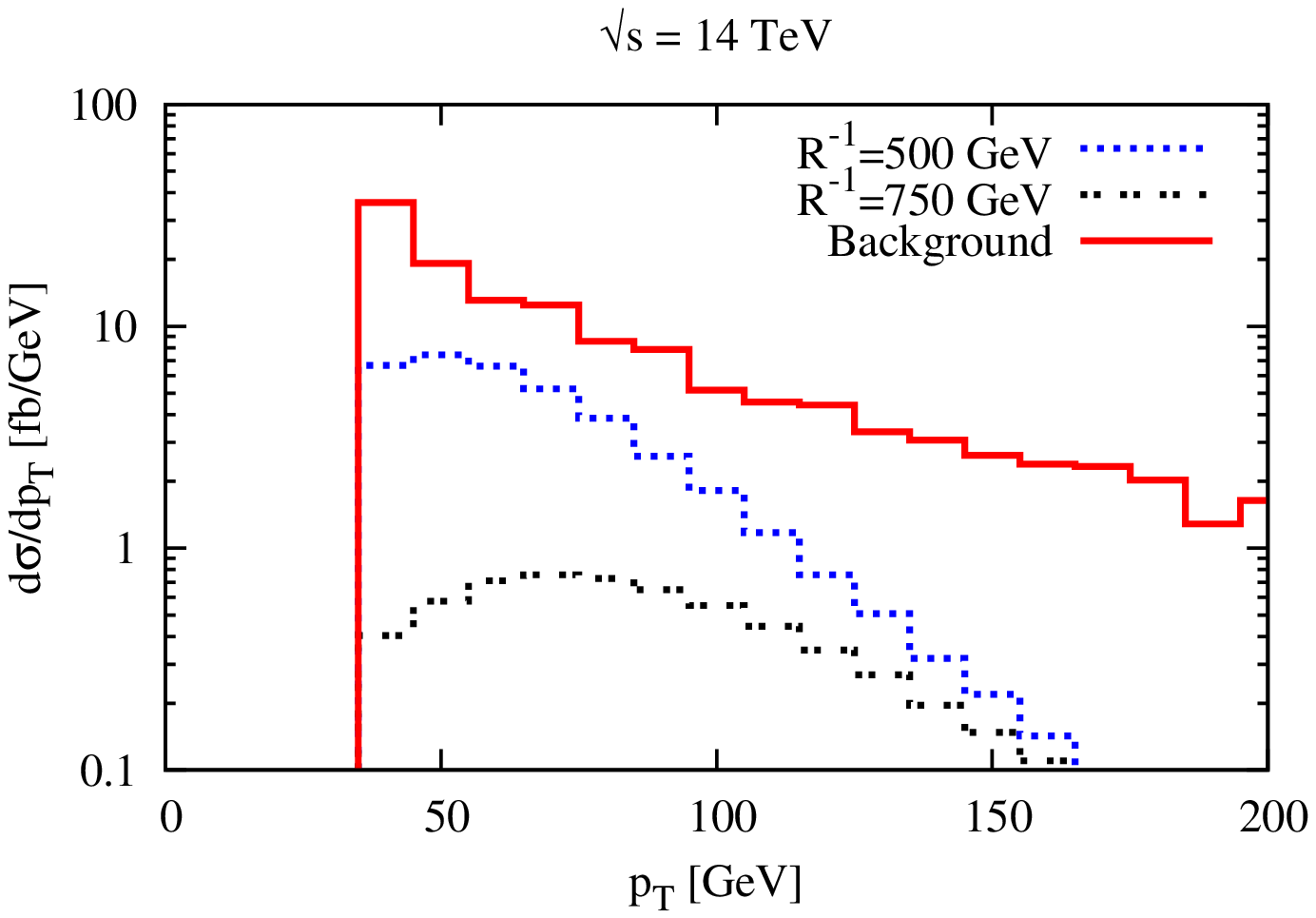,width=8cm,height=8cm,angle=-90}
\end{center}
\caption{Photon $p_T$ distributions after the {\em acceptance cuts} for both 
signal and the SM background at the LHC with $\sqrt s=$ 7 TeV (left), and 
14 TeV (right). } 
\label{fig:gammapt}
\end{figure}
\begin{figure}[h]
\begin{center}
\epsfig{file=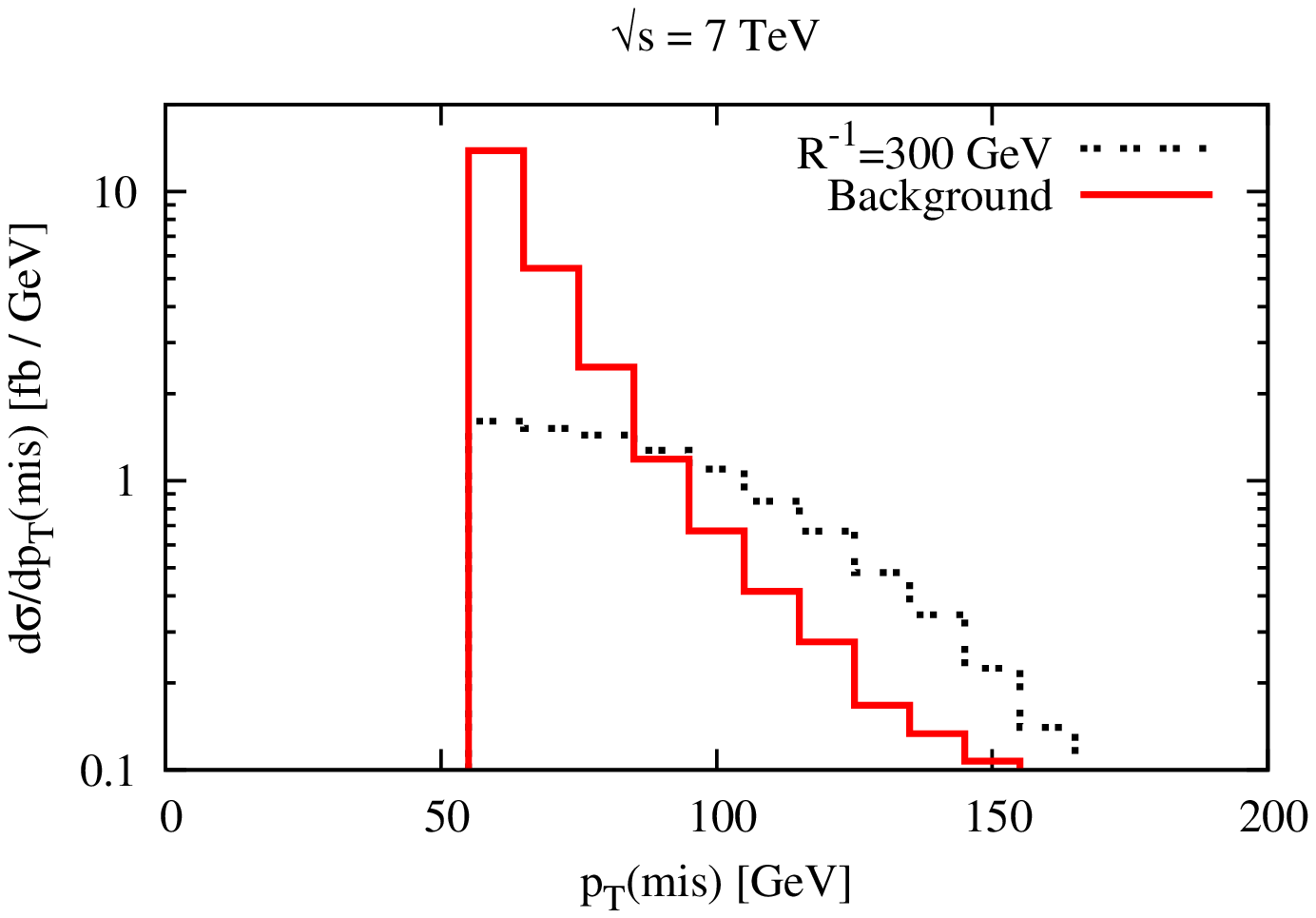,width=8cm,height=8cm,angle=-90}
\epsfig{file=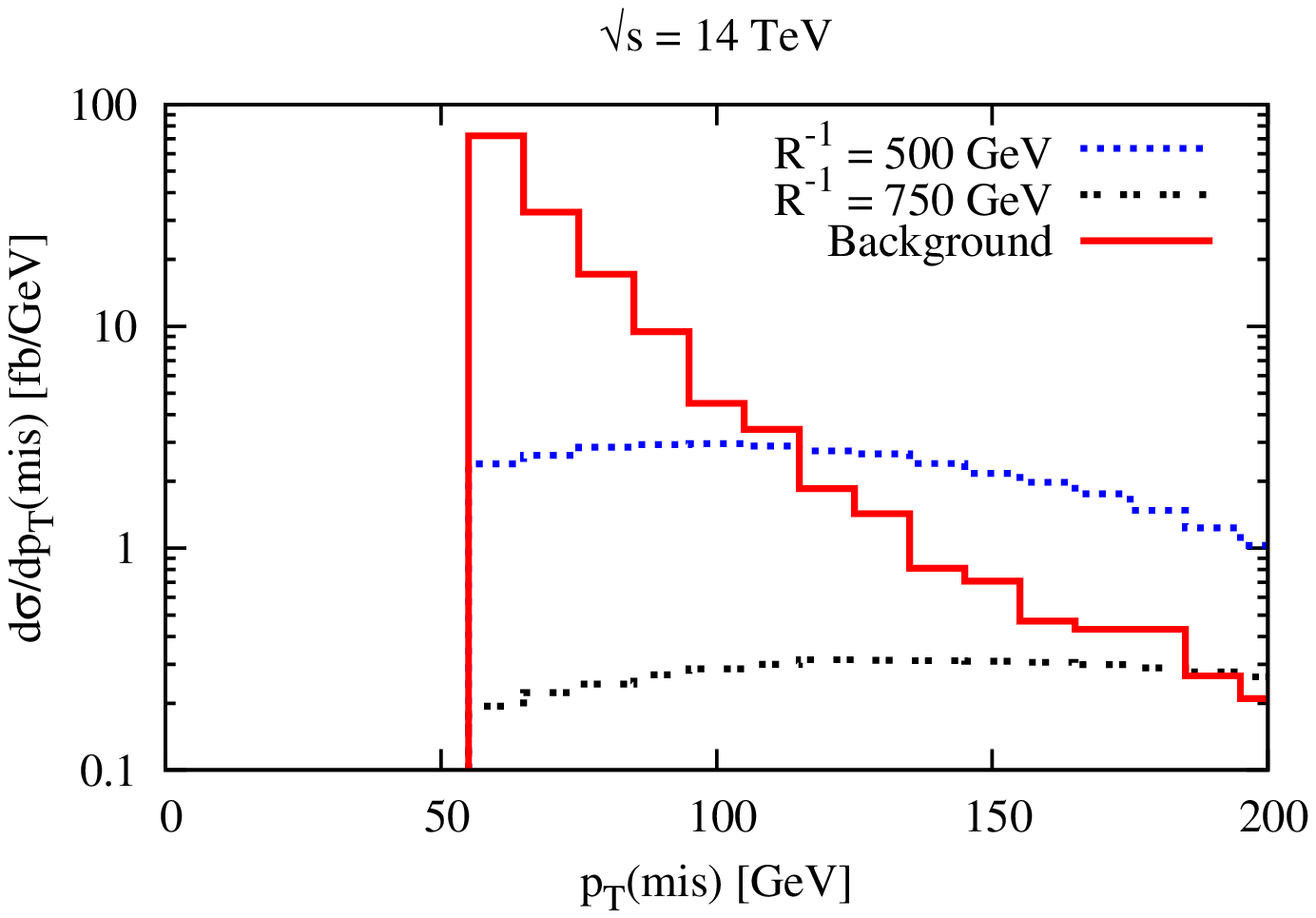,width=8cm,height=8cm,angle=-90}
\end{center}
\caption{Missing transverse momentum distributions after the 
acceptance cuts for both signal and the SM background at the 
LHC with $\sqrt s=$ 7 TeV (left), and 14 TeV (right). } 
\label{fig:mispt}
\end{figure}
\begin{figure}[t]
\begin{center}
\epsfig{file=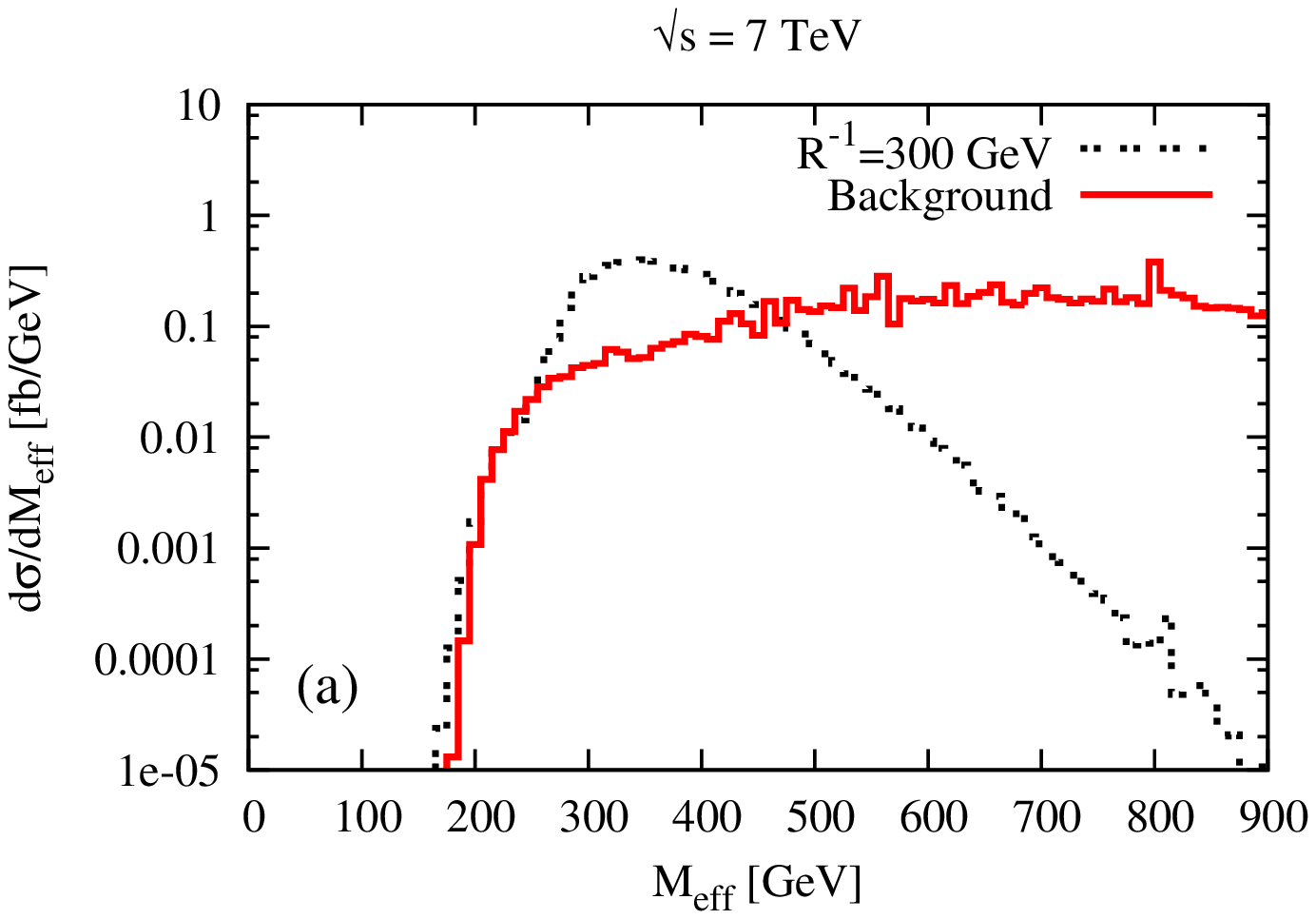,width=8cm,height=8cm,angle=-90}
\epsfig{file=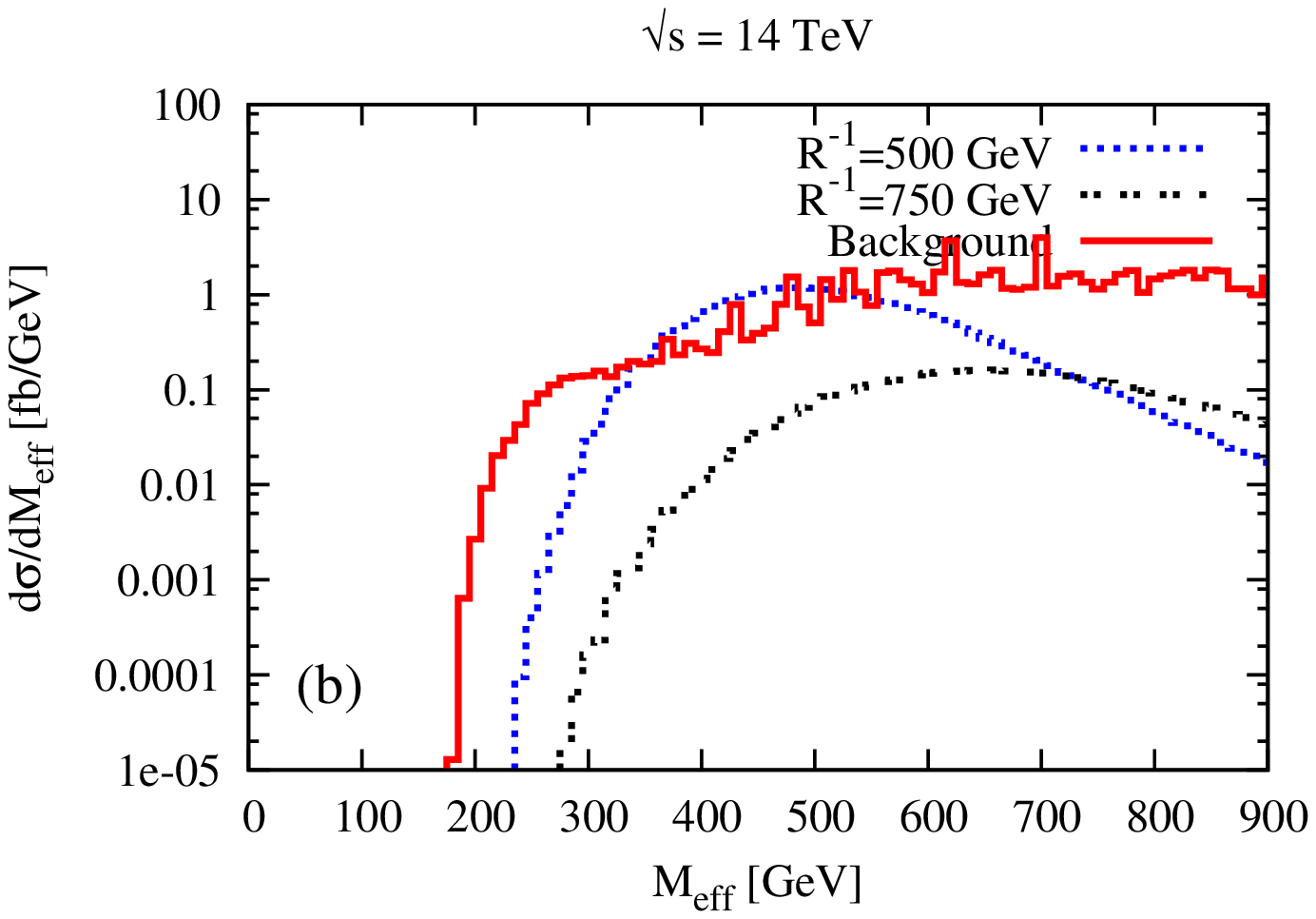,width=8cm,height=8cm,angle=-90}
\end{center}
\caption{Effective mass distributions after the acceptance cuts for 
both signal and the SM background at the LHC with $\sqrt s=$ 7 TeV (left), 
and 14 TeV (right). } 
\label{fig:Meff}
\end{figure}

Several SM processes constitute 
potential backgrounds for the 
signal of Eq.(\ref{signal}) and we now discuss the dominant 
ones in succession.
\begin{itemize}
\item An irreducible background arises from the production of a
  $Z$-boson in association with a {\em photon} and multiple jets.  The
  $Z$-boson decays invisibly and gives rise to the missing transverse
  energy signature: 
 \beq 
    pp ~\to~ Z+\gamma+n \mbox{--jets} ~\to~ \nu \bar \nu 
        +\gamma+n \mbox{--jets} 
\eeq 
We use the {\sc Alpgen} \cite{Mangano:2002ea}
generator to estimate the background contribution originating from the
above SM process.  Although the total cross section for this
  process is very large, the imposition of sufficiently strong $p_T$
  and rapidity requirements on the photon and the jets serves to
  suppress it strongly. In particular, the radiation of every
  additional hard (and well-separated) jet would, typically, cost and
  additional factor of $\alpha_s$.  However, since ours will not be a
  very sophisticated analysis, it is quite conceivable that we might
  underestimate the background, especially where jet reconstruction is
  concerned. To compensate for this, we will include even ostensible
  3--jet events in the background while requiring $n \geq 4$ for the
  background.  

\item The production of $W^\pm$ in association with a photon
 and multiple jets can also be a possible
  source of background,if the $W^\pm$ decays
  leptonically and the charged lepton is missed somehow. To be specific, we
    consider it to be undetectable if it either falls outside the
    rapidity coverage ($|\eta|\ge 2.5$) or if it is too soft ($p_T\le
    10$ GeV) or if it lies too close to any of the jets. In this
  case, the neutrino and the  missing
  lepton together give rise to the missing transverse momentum.
%
%
  This background
  too we estimate using {\sc Alpgen}.  Given the fact that the $W$ has a
  substantial mass and that it is produced with relatively low
  rapidity, it stands to reason that the charged lepton would, most
  often, be well within the detector and also have sufficient $p_T$ to
  be detectable. Consequently, the
  probability of missing the charged lepton is small, and this
  background would be suppressed considerably, Indeed, this is borne
  out by actual computation.  

\item Significant background contribution can come from the 
production of a {\em photon} in association with 
four or more jets. 
\beq
pp~\to~ \gamma+nj~~~~~~~~~~{\rm with}~~~~~n\ge 4 \nonumber
\eeq
In this case, there is no real source of missing transverse 
momentum. However,  mis-measurement of the $p_T$ of 
jets and photon can lead to some amount of missing transverse 
momentum.
Since the cross section for the aforementioned process is large,
this process, in principle, could
contribute significantly to the background.

\item Top anti-top ($t\bar t$) pair production in association 
with a photon is another source:
\beq
pp~\to~ t \bar t~ \gamma \ .
\eeq
Hadronic decay of the $t\bar t$ pair 
could lead to a final state comprising of, say,  6 jets alongwith 
a photon. The missing $p_T$ would result
from the mis-measurement of the jets 
and photon momentum. Similarly, the semi-leptonic decay of 
the $t\bar t$ pairs 
($pp\to t\bar t \gamma\to 4 q +\gamma+l+\nu$) also 
contributes to the background if the  charged lepton is
missed. In this case, the neutrino momentum would add to the 
contribution from the mismeasurement to yield the total missing $p_T$.

\item Single top (in association with a quark or a $W$) production, although 
an electroweak process, has a production cross section at the LHC that is 
quite comparable to that for $t \bar t$. Radiating off a photon (as in 
$p p \to t + \gamma + X$ with $X$ not being a top) would, naively, 
result in a background that 
is suppressed compared to that from $t \bar t \gamma$ by a factor 
$ \sim \sigma(t + X) / \sigma(t \bar t) $. However, note that, for a very large 
fraction of these events, $X$ is a quark and produced preferentially with a
low $p_T$ and/or high rapidity. Such and other considerations turn 
this background quite subdominant.

\item Also to be considered are events such as 
\[
   p p \to W^\pm + n \mbox{--jets} \to e^\pm + \nu + n \mbox{--jets}.
\]
If the $e^\pm$ does not leave a track in the detector, then a hit in 
the electromagnetic calorimeter would qualify it as a photon, thereby 
mimicking the signal of Eq.(\ref{signal}). Pending a full detector 
simulation, an accurate estimate of this background is not possible. 
However, the probability 
of an $e^\pm$ of $p_T > 10 \gev$ not leaving a track is very 
low~\cite{CMS_elec_recons}. As a result, this background 
is expected to be suppressed in comparison to that from 
($ W^\pm + \gamma + n \mbox{--jets}$)--production despite the 
latter appearing only at a higher order of perturbation theory.

\item Last, but not the least, is the pure QCD background, namely 
just $n$-jet production, with one jet faking a photon. With the rates 
for $n$-jet production being quite large, it is conceivable that this 
background could be substantial. Again, an accurate estimation would need 
a full detector simulation. However, as previous studies have shown, 
for sufficiently high values of the minimum $p_T$ required, the 
probability of photon faking is $\sim 10^{-3}$--$10^{-4}$, and, once again,
this background turns out to be smaller than that from 
$\gamma +n$--jets.

\end{itemize}
\begin{figure}[t]
\begin{center}
\epsfig{file=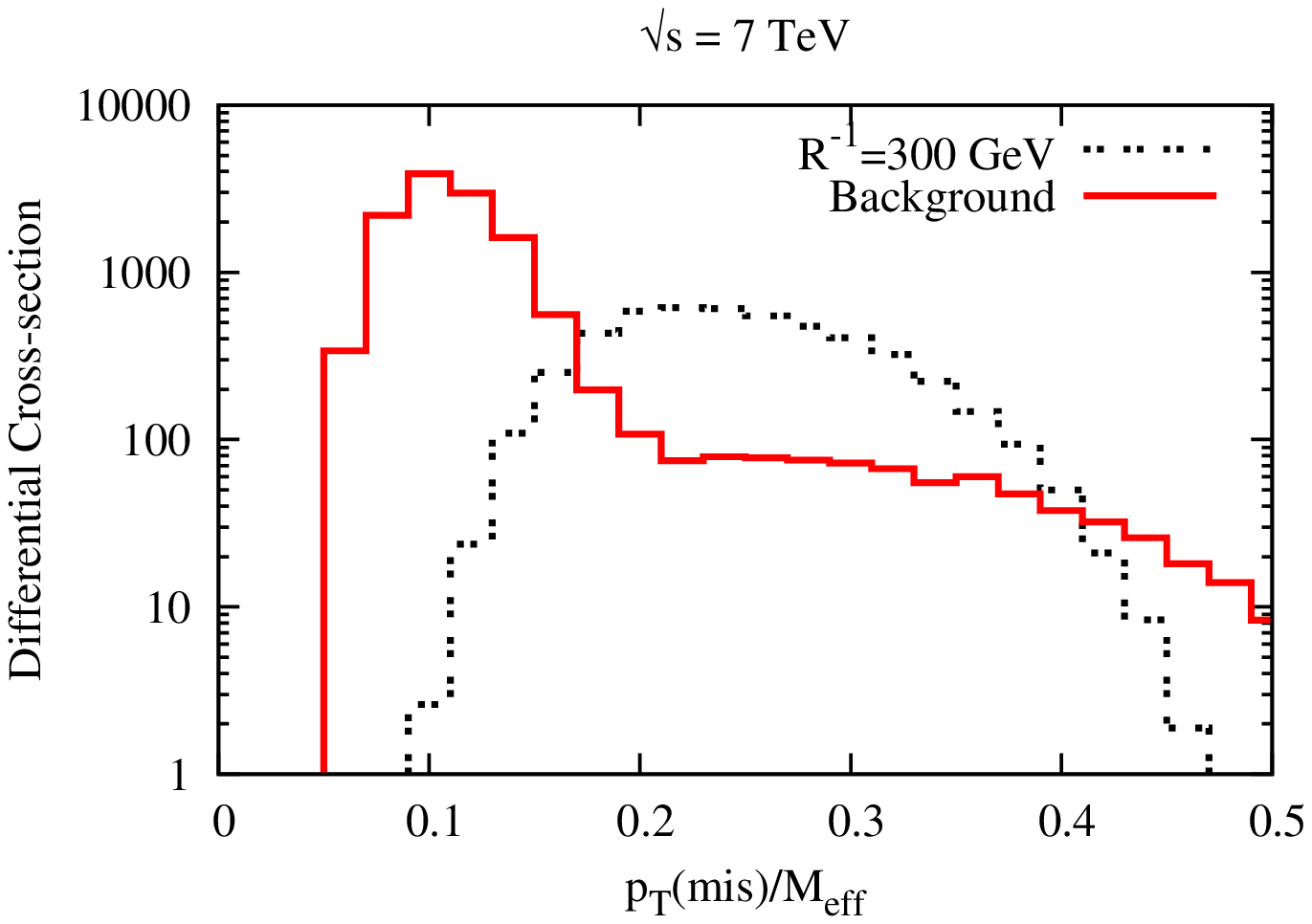,width=8cm,height=8cm,angle=-90}
\epsfig{file=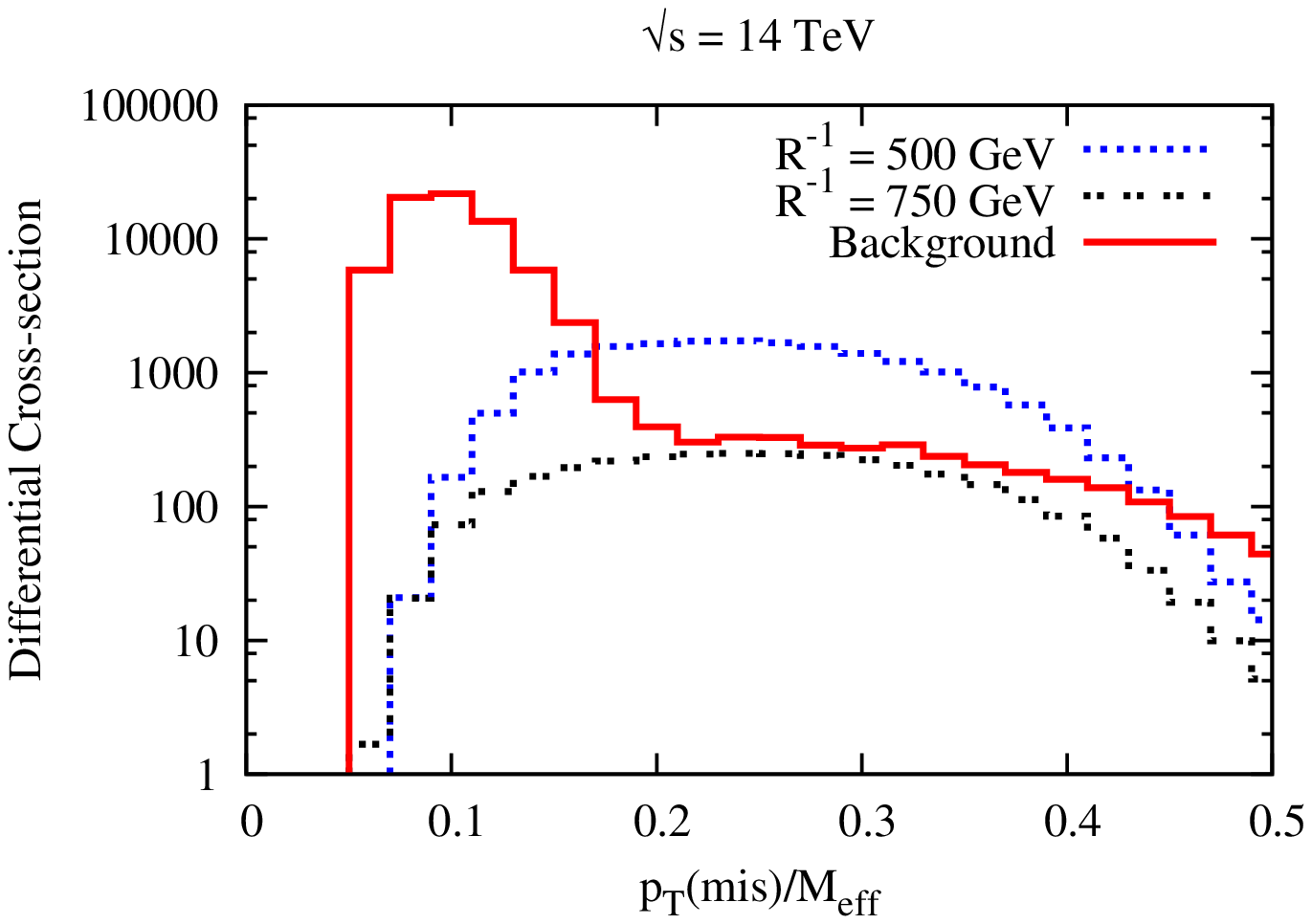,width=8cm,height=8cm,angle=-90}
\end{center}
\caption{Distributions in the 
ratio $\ptmiss /M_{\rm eff}$ after the 
acceptance cuts for both signal and the SM background at the 
LHC with $\sqrt s=$ 7 TeV (left) and 14 TeV (right). } 
\label{fig:ratio}
\end{figure}
At this stage, we are equipped enough to develop a systematic
methodology of suppressing the SM backgrounds without drastically
reducing effecting the signal.  A fruitful perusal of such a
methodology requires that we carefully examine and compare the phase
space distributions of different kinematic variables for signal as
well as backgrounds discussed above. However, before we embark on the
mission to suppress the aforementioned backgrounds, it is important to
list a set of basic requirements for jets and photons to be visible at
the detector.  It should be appreciated that any realistic detector
has only a finite resolution; this applies to both energy/transverse
momentum measurements as well as the determination of the angle of
motion. For our purpose, the latter effect can be safely
neglected\footnote{The angular resolution is, generically, far
superior to the energy/momentum resolutions and too fine to be of any
consequence at the level of sophistication of this analysis.} and we
simulate the former by smearing the energy with Gaussian
functions. The energy resolution function receives contributions from
many sources and are, in general, a function of the detector
coordinates. We, though, choose to simplify the task by assuming a
flat resolution function equating it to the worst applicable for our
range of interest, namely,
\bea
\frac{\Delta E}{E}&=&\frac{a}{\sqrt {E/{\rm GeV}}}\oplus b,
\eea
where, $a_\ell=5\%,~b_\ell=0.55\% $ and $ a_j=100\%, b_j=5\%$, and 
$\oplus $ denotes a sum in quadrature. Keeping in mind 
the LHC environment as well as the detector configurations, 
we demand that, to be visible, a jet or a photon must have an 
adequately large transverse momentum and they are well inside 
the rapidity coverage of the detector, 
namely,
\beq
p_T^{j} > 30 \gev \ , \qquad p_T^{\gamma} > 30 \gev \ ,
\label{cut:pT}
\eeq
\beq
|\eta_{j}| \leq 2.5 \ , \qquad |\eta_{\gamma} | \leq 2.5 \ .
\label{cut:eta}
\eeq
We demand that both photon and jets  
be well separated 
so that they can be identified as individual entities. 
To this end, we use the well-known cone algorithm 
defined in terms of a cone angle 
$\Delta R_{ij} \equiv \sqrt{\left (\Delta \phi_{ij}\right)^2 
+ \left(\Delta \eta_{ij}\right)^2} $, with 
$\Delta \phi $ and $ \Delta \eta $ being the azimuthal angular 
separation and rapidity difference between two particles.
Quantitatively, we impose
\beq
\Delta R_{\gamma \,j} > 0.4 ~{\rm and}~ \Delta R_{j \, j} > 0.7.
\label{cut:jj-iso}
\eeq
Furthermore, the event must be characterized by a 
minimum missing transverse momentum defined in terms of the 
total visible momentum, namely,
\beq
\not p_T \equiv \sqrt{ \bigg(\sum_{\rm vis.} p_x \bigg)^2 
                 + \bigg(\sum_{\rm vis.} p_y \bigg)^2 } \; > 50 \gev \ .
\label{cut:pt-mis}
\eeq 
It has been discussed already that for some of the SM 
backgrounds, the hard (parton level) process does not even 
have a source of missing energy. For example, 
the $4j$+$\gamma$ final state could potentially be associated 
with a missing transverse momentum only
on account of 
mismeasurements of the jets and photon energies. A minimum 
requirement of the missing transverse momentum 
keeps these backgrounds well under control. 
We  also impose 
a lower 
bound on the {\em jet-jet} and {\em jet-photon} invariant mass.
 \beq
M_{\gamma \,j} \ge 20~{\rm GeV} ~{\rm and}~  M_{j \, j} \ge 20~{\rm GeV}.
\label{cut:M}
\eeq
The last two cuts are aimed to avoid any collinear and soft singularities
associated with the emission of a photon/gluon from a quark, while estimating
the background processes, which we evaluate using LO matrix elements.
The requirements summarised  in 
Eqs. (\ref{cut:pT}--\ref{cut:M}) constitute our 
{\em acceptance cuts}.

With the set of acceptance cuts and detector resolution defined 
in the previous paragraph, we 
compute the signal and background cross-sections 
at the LHC  operating with $\sqrt{s} = 7 $ TeV and 14 TeV 
respectively and display them in Table~\ref{tab:cross}.
Clearly, the backgrounds
are still large compared to the signal.  The dominant SM background
contribution arises from the production of a photon in association
with jets\footnote{Indeed, the dominance is strong enough to render
inefficient the use of additional cuts such as ones on $|M_{jj} -
M_{Z/W}|$ to eliminate resonant $Z/W$ production.}.  In order to
enhance the signal to background ratio, we study distributions of
different kinematic observables.

\begin{table}[h]
\begin{center}
\begin{tabular}{|c|c||c|c|c|}
\hline \hline
\multicolumn{5}{|c|}{Cross-section in fb}\\
\hline\hline
\multicolumn{2}{|c||}{$\sqrt s=7$ TeV}  & \multicolumn{3}{|c|}{$\sqrt s=14$ TeV}
\\\hline
$R^{-1}$ [GeV]& Background & \multicolumn{2}{|c|}{$R^{-1}$ [GeV]} & Background \\
300 & cross-section & 500 & 750 & cross-section \\\hline
98.6 & 252.5 & 376.3 & 62.1 & 1478.1 \\\hline\hline 
\end{tabular}
\end{center}
\caption{Signal and SM background cross-sections (in fb) after the acceptance 
cuts for different values of $\sqrt s$ and $R^{-1}$.}  
\label{tab:cross}
\end{table}

In Fig.~\ref{fig:jetpt}, we 
display the $p_T$ distributions of the signal 
(top panels) and background (bottom panels) jets after 
ordering them according to their $p_T$ 
($p_T^{j_1}>p_T^{j_2}>p_T^{j_3}>p_T^{j_4}$). 
From the shape of the $p_T$ distributions in 
Fig.~\ref{fig:jetpt} it is very obvious that the signal 
jets are relatively softer
 than the background jets. 
The  former result from the decay 
of excited gluons ($G_\mu^{(1,0)}$) and quarks 
($Q_-^{(1,0)}~{\rm and}~ Q_-^{(1,0)}$) into other excited 
electroweak gauge bosons and spinless adjoints. 
With the relative mass splitting between the excited 
quarks/gluons and the weak gauge bosons and spinless adjoints 
being small, it is quite obvious that, in the rest frame of the 
primary produced particle, the daughter jets would 
carry only a small fraction of its mass as momenta. This, in 
turn, translates to relatively small transverse momenta for 
them. 
This characteristic of the signal could, in
principle, be exploited to enhance the signal to background
ratio. 
However, while an upper bound on the jet $p_T$ would suppress 
the SM backgrounds, it is important to
notice that (see Fig.~\ref{fig:jetpt}) the hardness of signal jet
$p_T$s depends on the center-of-mass ($\sqrt {s_{pp}}$) energy of the
collider as well as on the compactification radius $R$.
Thus, to suppress the SM background without
reducing the signal, such an upper bound (if any), has to be designed
keeping in mind both both $\sqrt {s_{pp}}$ and $R^{-1}$. 
Similar is the situation with  the photon $p_T$ distribution 
(see Fig.~\ref{fig:gammapt}). 
In Fig.~\ref{fig:gammapt}, we 
show the $p_T$ distributions for the signal and background photons
for $\sqrt{s} = 7 $ TeV (left panel) and 14 TeV (right panel)
of the LHC center of mass energy. 
Here too, a 
$\sqrt {s_{pp}}$ and $R^{-1}$ dependent upper bound on the photon
transverse momentum would improve the signal to noise ratio. 
However, in absence of any information about
$R^{-1}$, it is extremely challenging to introduce such a
cut. Therefore, in our analysis, we do not use any further $p_T$ cuts.

In Fig.~\ref{fig:mispt}, we display the missing transverse momentum
distribution for the signal and background for two values of LHC
center of mass energies. The background is peaked at a relatively low
$\ptmiss$. This is a consequence of the fact that with the {\em
acceptance cuts} (Eq.~(\ref{cut:pT}--\ref{cut:M})), the dominant SM
background contribution arises from the {\em $\gamma+n$-jets}
production.  Since, for this process, a missing
transverse momentum can arise only from 
mis-measurement, this contribution
can be suppressed significantly by introducing  a harder $\ptmiss$
cut. However, this would also reduce the signal 
simultaneously (see Fig.~\ref{fig:mispt}) 
and, hence, we
do not impose any further missing transverse momentum cut in our
analysis.  

Another variable that is often used for such purposes is
the effective mass. Defined as the scalar sum of the transverse
momenta of all the visible particles, as well as the total missing
transverse momentum, it can be expressed, in our case, through
\beq
M_{\rm eff}=\sum_{j}p_T^{j}+p_T^\gamma+ \ptmiss \ .
\label{meff}
\eeq

In Fig.~\ref{fig:Meff}, we show the $M_{\rm eff}$ distributions of the
signal and background at the LHC with $\sqrt{s} = 7 $ TeV (left panel)
and 14 TeV (right panel) respectively. 
Expectedly, the distribution 
is flatter for larger $R^{-1}$ (see Fig.~\ref{fig:Meff}$b$).
And, while it may seem that, for $\sqrt{s} = 7 \tev$, the signal
distribution does rise above the background, note that this is true 
only for a relatively low value of $R^{-1}$. Furthermore, with
the peak position being a strong function of $R^{-1}$,  a cut on $M_{\rm eff}$ 
would be effective only if it is designed accordingly. In other words, such 
a cut would not be very useful tool in search strategies.

Finally, we consider the ratio $\ptmiss /M_{\rm eff}$, and
in Fig.~\ref{fig:ratio}, present the distributions in the same. 
The background peaks around $\ptmiss /M_{\rm eff} \sim 0.1$
and it is obvious that it would be reduced significantly 
if a lower bound on this ratio is imposed 
this ratio. 
To be specific 
In our analysis, we require that 
\beq
   \frac{\ptmiss}{M_{\rm eff}} \geq 0.2
   \label{ptm_meff_ratio}
\eeq
Therefore, our final event {\em selection criteria} consists of 
the {\em acceptance cuts} (viz. Eq.~(\ref{cut:pT}--\ref{cut:M}))
alongwith  Eq.(\ref{ptm_meff_ratio}).
In Table~\ref{tab:cross_sec}, 
we summarize
the signal and the SM background cross-sections, for two 
operative energies of the LHC, after the imposition of all of these cuts.
It is very evident that more 
than $5\sigma$ discovery for $R^{-1}=500$ GeV is possible with 
the integrated luminosity of $2{\rm fb}^{-1}$ at the LHC 
with $\sqrt s=7$ TeV. On the other hand, if the LHC reaches
 $\sqrt{s} = 14$ TeV, 
 we will be able to probe 
$R^{-1}$ upto 1000 GeV with an integrated luminosity of 
20 ${\rm fb}^{-1}$.
\begin{table}[!h]

\begin{center}

\begin{tabular}{|c|c|c||c|c|c|}
\hline \hline
\multicolumn{6}{|c|}{Cross-section in fb}\\
\hline\hline
\multicolumn{3}{|c||}{$\sqrt s=7$ TeV}  & \multicolumn{3}{|c|}{$\sqrt s=14$ TeV}
\\\hline
Background & $R^{-1}$ & Signal  & Background & $R^{-1}$ & Signal \\ 
in fb & GeV & [fb]  &[fb] & GeV & [fb] 
\\\hline\hline
 & 300 & 73.1  &  & 500 & 265.1 \\
12.12 &  &   & 47.88 & 750 & 43.9 \\
 & 500 & 17.5  &  & 1000 & 9.4 \\\hline\hline

\end{tabular}
\end{center}

\caption{Signal and SM background cross-sections (in fb) after the {\em selection cuts} for different values of $\sqrt s$ and $R^{-1}$.}  

\label{tab:cross_sec}
\end{table}

\subsection{Possibility of mass measurement/ parameter determination}

\begin{figure}[t]
\begin{center}
\epsfig{file=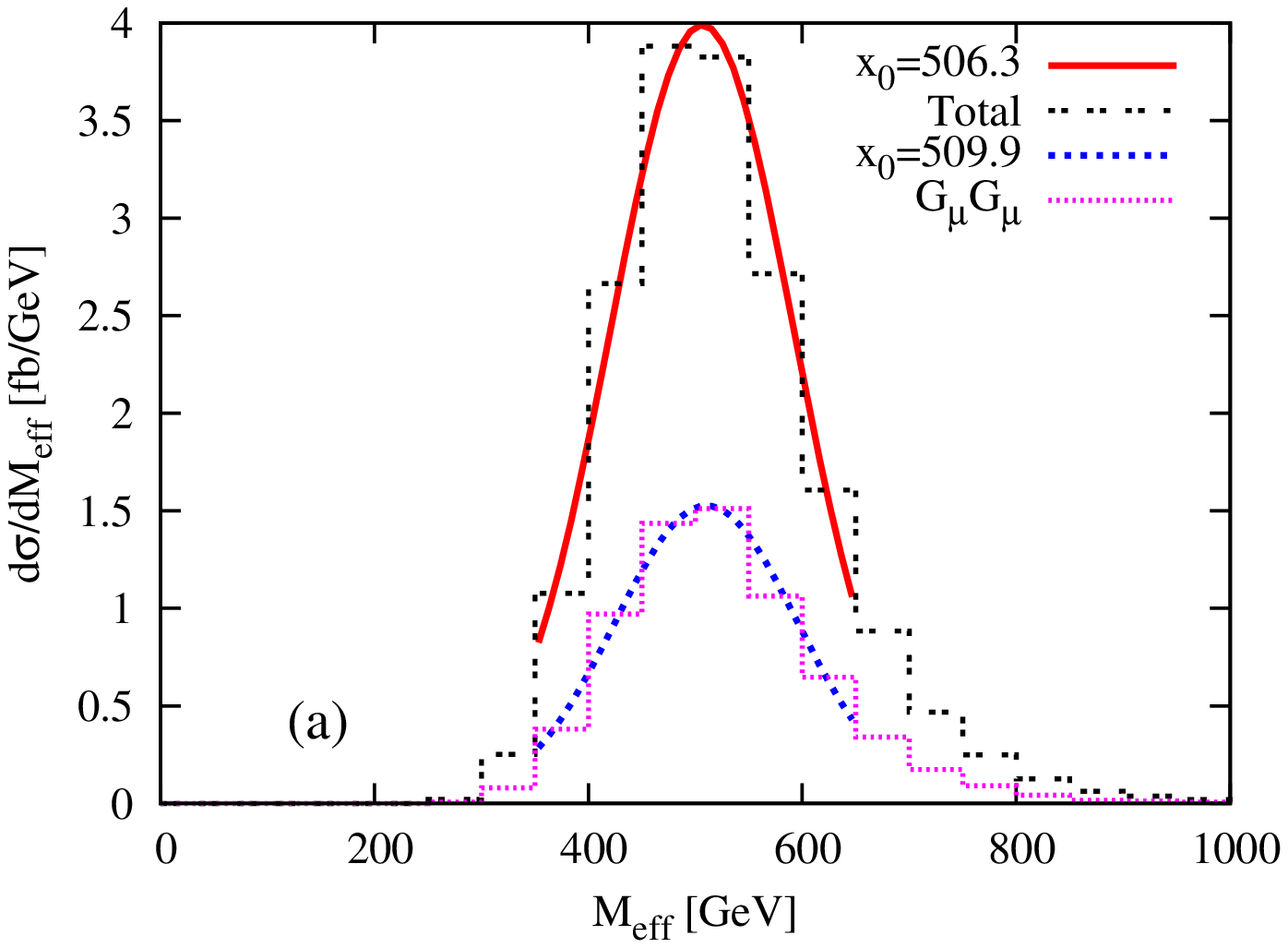,width=7cm,height=7cm,angle=-90}
\epsfig{file=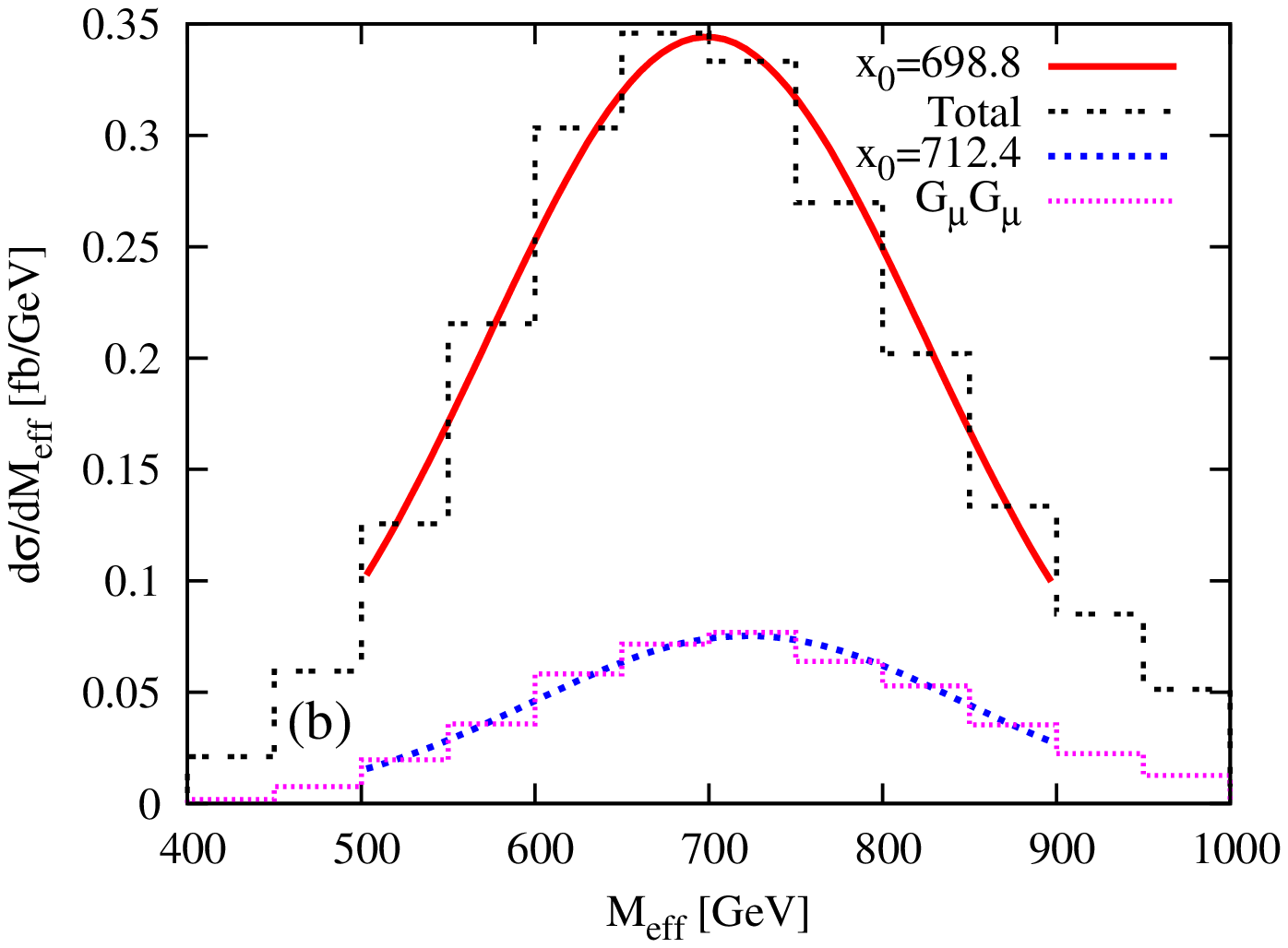,width=7cm,height=7cm,angle=-90}
\end{center}
\caption{The total $M_{eff}$ distributions and contribution from 
$G_\mu G_\mu$ production for $R^{-1}$= 500 GeV (left panel) and 750 GeV 
(right panel) with $M_s = 10 R^{-1}$  at the LHC with center-of-mass energy 
14 TeV. $x_0$ stands for the peak position of the respective fitted Gaussian 
distribution.}
\label{fig:justification}
\end{figure}

The search strategy, presented above, does not depend on looking for
any bump or edges. Consequently, a direct determination of masses or
mass differences is not straightforward.  Rather, the analysis is of
the number counting type and, thus, depends crucially on an accurate
estimation of the SM backgrounds. In particular, both signal and
background comprise of multiple particle production of varied type, as
well as cascade decay chains.  And last but not the least, is the fact
that the whole analysis pertains to a hadron collider, where any
theoretical prediction of number of events/cross-section is to be
accepted with the uncertainties of parton distribution, choice of
scales etc. Interplay of these two effects prevents a determination of
masses from the cross-section measurement itself.
 
However, a careful look at the $M_{\rm eff}$ (defined in Eq. (\ref{meff}))
distributions in Fig.~\ref{fig:Meff}, leads us to a possible
correlation between the peak position ($x_0$) of such  distributions
with respective $R^{-1}$ values.  

Naively, one would expect this correlation to become stronger once the
cut of Eq.(\ref{ptm_meff_ratio}) is imposed, and this, indeed, turns
out to be the case. However, before we attempt to establish this
correlation, we turn to the question of multiple particle production
channels contributing to the signal, for, despite the KK spectrum
being relatively degenerate, multiple peaks would still be expected.
To examine this, we consider, individually, each of the major
production channels contributing to the signal. After background
removal, the excess is then fitted with a Gaussian. Remarkably, the
peak positions are not too different. This is illustrated in
Figs.~\ref{fig:justification}, wherein we present the $M_{eff}$
distributions for two different values of $R^{-1}$ (500 GeV and 750
GeV) wherein both the total signal (adding up all the sub-processes)
and the contribution from $G_\mu G_\mu$ production alone have been
presented separately. Note that $G_\mu G_\mu$ corresponds to the
heaviest masses corresponding to a particular choice of parameters,
and thus, represents the maximum deviation from the overall sum. This
small difference is, of course, reflective of the relative degeneracy
of the spectrum. Furthermore, the absence of any discernible
thresholds (or the existence of mutiple peaks) in the distribution for
the entire signal, is a consequence of the twin facts that the
separations between the peaks is much smaller than the widths of the
individual distributions and that no dominant peak exists below the
$G_\mu G_\mu$ one.

A few points need to be made at this juncture. The relative importance
of the $G_\mu G_\mu$ channel might seem unwarranted in view of the
fact that this does not represent the dominant production process (see
Fig.\ref{fig:cross}). However, note that one of the two dominant mass
splittings in the theory is that between the $G_\mu$ and
$Q_\pm$. Consequently, the jet from the $G_\mu \to Q_\pm$ decay is a
energetic one. Thus, the requirement of four jets with $p_T > 30 \gev$
preferentially selects $G_\mu G_\mu$ events, thereby according it far
greater importance. The relatively large separation between the $G_\mu
G_\mu$ peak and the overall peak for a larger $R^{-1}$ value (compare
Figs.~\ref{fig:justification}{ a \& b}) is also a testament to
this, for a larger $R^{-1}$ implies a larger absolute split between
the different level--(1,0) fields, leading to harder jets all
around. However, even for $R^{-1} = 1 \tev$, the difference is
comparatively small when compared to experimental resolutions. The
final point relates to the fact that peak in $M_{\rm eff}$ is closer
to the masses of the particle produced rather than twice the
mass. This is not unexpected, given the fact that the spectrum is so
degenerate (and, consequently, softer jets) and that the two
contributions (emanating from the two $B_H^{(1,0)}$) to missing $p_T$
cancel each other to a significant extent. Indeed, the situation
is somewhat analogous to that in Ref.\cite{biswarup_kirtiman_satya}.

\begin{figure}[h]
\begin{center}
\epsfig{file=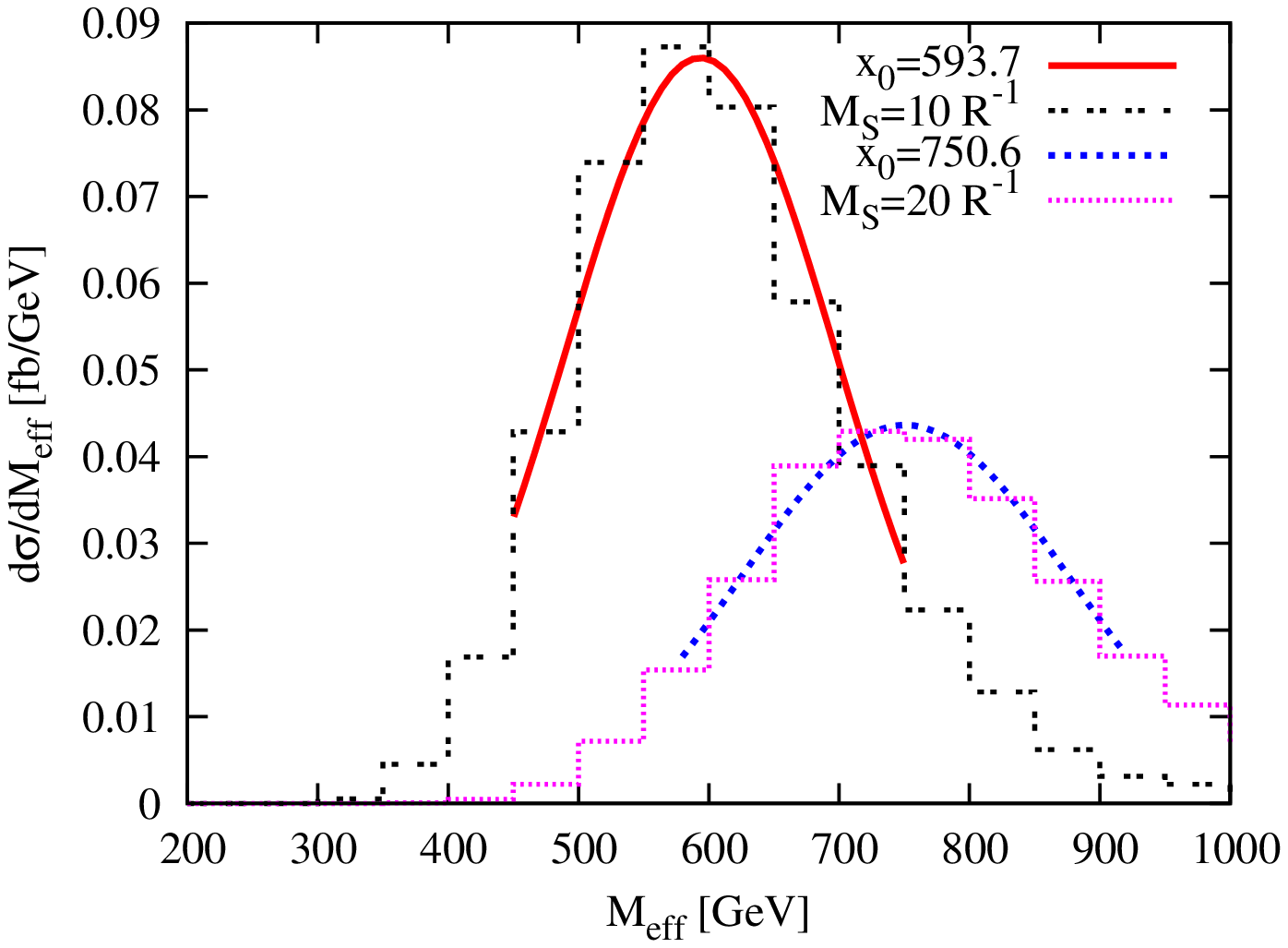,width=5cm,height=5cm,angle=-90}
\epsfig{file=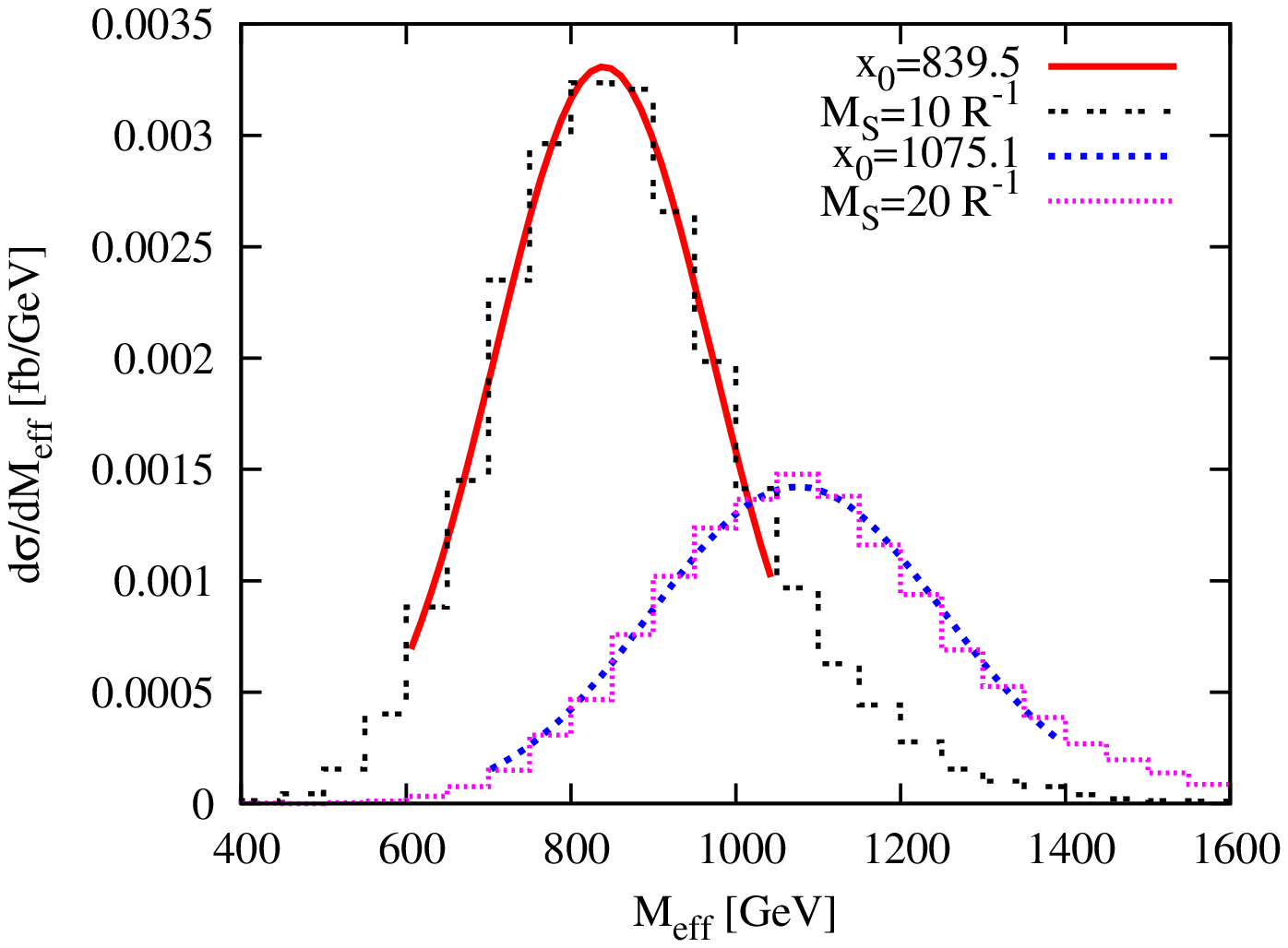,width=5cm,height=5cm,angle=-90}
\epsfig{file=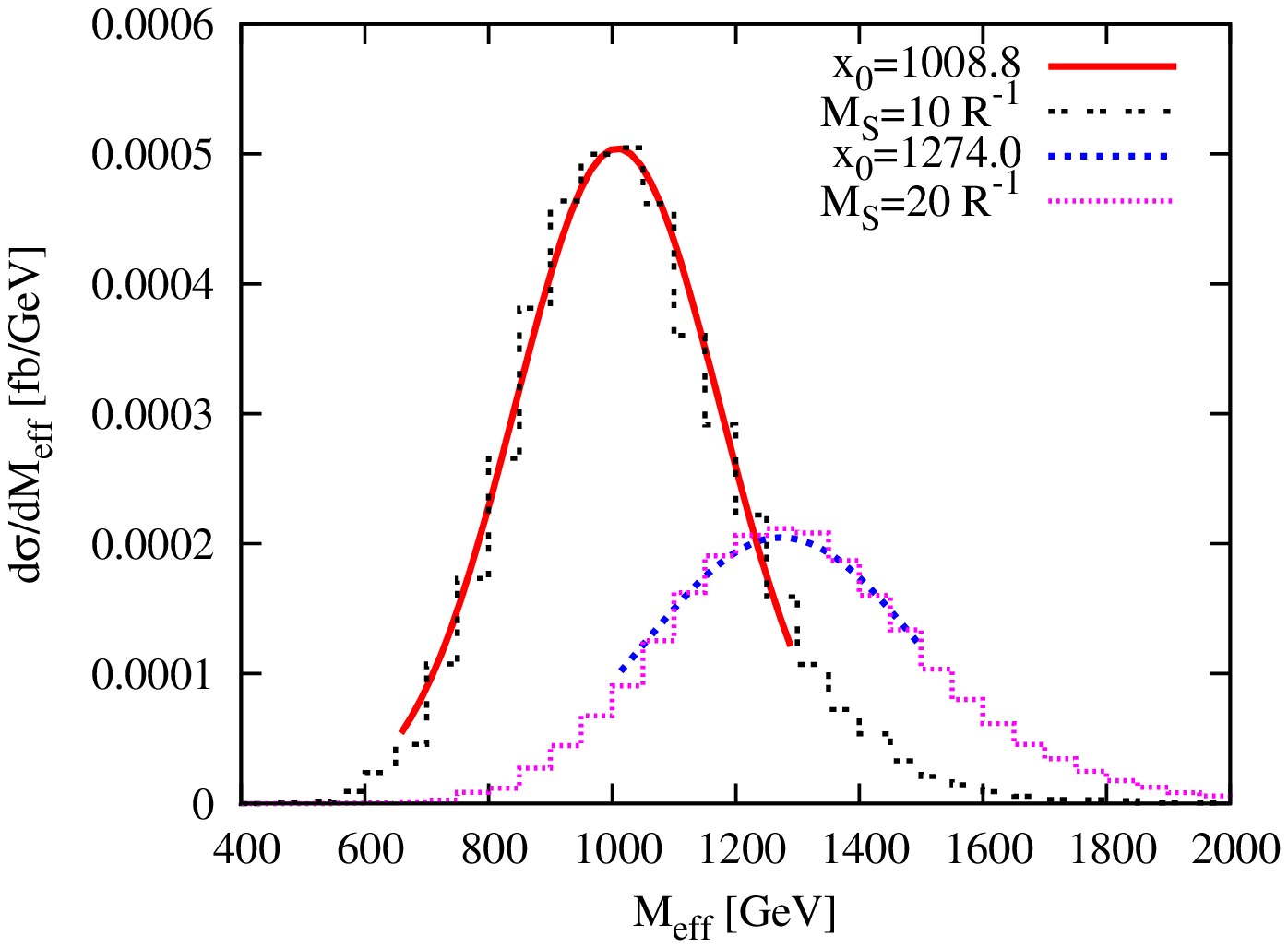,width=5cm,height=5cm,angle=-90}
\end{center}
\caption{The $M_{eff}$ distributions (calculated using only the $G_\mu G_\mu$
sub-process) for different values of $R^{-1} (= 600, 900, 1100$ GeV)
and cut-off $M_s (= 10 R^{-1}$ and $20 R^{-1}$) at the LHC with
center-of-mass energy 14 TeV.}
\label{fig:fit}
\end{figure}

It ought to be remembered that, for a given $R^{-1}$, the 
quantum corrections to the masses (and, hence, the splitting between 
them) grows with the value of the cut-off $M_s$. With the loss 
of degeneracy, the position of the $M_{\rm eff}$ peak would 
move upwards. This is amply demonstrated by Fig.\ref{fig:fit}, wherein
we present the $M_{\rm eff}$ distribution for the signal (on imposition
of all the selection cuts) alongwith the Gaussian fit. We have chosen to 
use only the $G_\mu G_\mu$ production for reasons explained above.

\begin{figure}[t]
\begin{center}
\epsfig{file=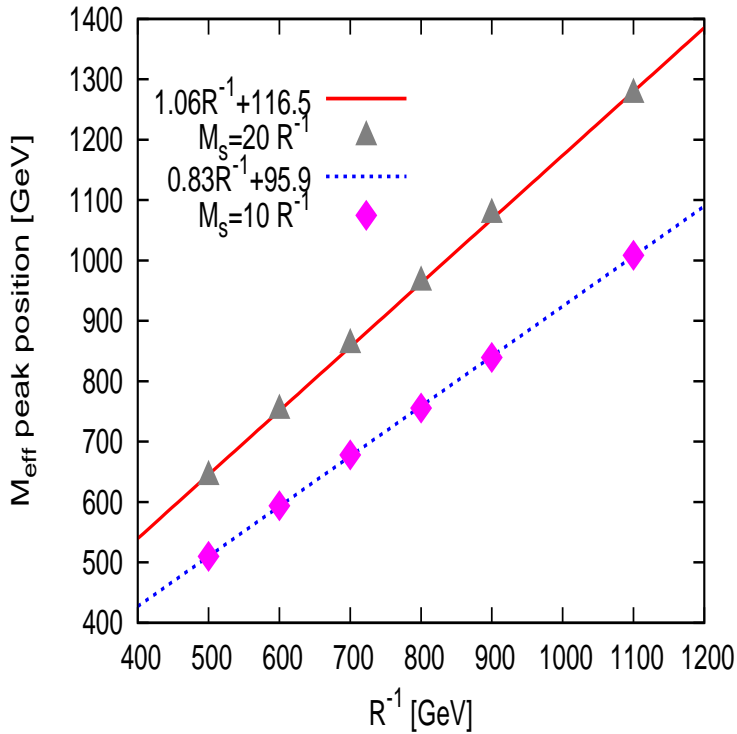,width=10cm,height=10cm,angle=-90}
\end{center}
\caption{Correlation between the peak of $M_{eff}$ distributions and
$R^{-1}$.}
\label{fig:correlation}
\end{figure}

In Fig.\ref{fig:correlation}, we exhibit the relation between $R^{-1}$
and the fitted peak position (denoted $x_0$). The linear 
dependence is understandable given the dependence of the masses on the
compactification radius. As could already be expected from a study of
Fig.\ref{fig:fit}, the exact linear relation (i.e., the coefficients)
does depend on the cutoff scale. Given this, it is obvious that an
accurate extraction of $R^{-1}$ (equivalently, the common mass scale)
is not possible from this measurement alone. However, given that $M_s
R$ cannot be too large, the inaccuracy, as suggested by
Fig.~\ref{fig:correlation}, is perhaps not too large for a first
estimate. An unique determination is possible only with further
experimental data. This could come about in a variety of ways. For
example, an accurate measurement of the signal cross section would
provide us with supplementary information about the mass
scale\footnote{While both the position of the peak and its height
depend on $R^{-1}$ as well as $M_s \, R$, the dependences are
different.}. Similarly, the discovery of any substructure in the
$M_{\rm eff}$ distribution would hint at (some of) the mass
differences. And, finally, the use of other final states would provide
additional information. However, given that ours is only a preliminary 
study, we desist from a more complete examination of this issue, 
for that
demands a detailed study inclusive of 
higher order effects on the one hand and 
detector simulation on the other.



\section{Summary and Conclusion}
To summarise, we have investigated the possibility of discovering the
Universal Extra Dimensional Model with 2 extra space-like dimensions
at the LHC.  We have considered the production of strongly
interacting $(1,0)$--mode particles, namely the KK-excitations of
gluons and quarks of first two generations. At the LHC, the total
production cross-section of such strongly interacting particles is
quite large. Once produced, 
these decay to
SM gluons and quarks along with $(1,0)$--mode EW gauge bosons/adjoint
scalars. Further decays of these gauge bosons and scalars will produce
leptons and photons.  
Of particular interest is the fact that the $B_\mu
^{(1,0)}$ (KK-excitation of the $U(1)$ gauge boson) decays into a
photon and $B_H ^{(1,0)}$, the latter being the lightest
KK-particle. In this article, we concentrated on a final state
comprising of multiple jets along with a single hard photon and
missing transverse energy (due to the production of a pair of $B_H
^{(1,0)}$ as the end products of the decay chains ). There are several Standard Model
backgrounds which can mimic similar final states in the detector.  We
have estimated cross-sections for the signal processes using {\sc
  CalcHep}~\cite{calchep} whereas
 the SM background processes have been
estimated using {\sc Alpgen}~\cite{Mangano:2002ea} and {\sc
  MadGraph}\cite{madgraph}.  While several different processes
contribute to the background, the dominant source is the SM process
$pp\to Z\gamma+n$-jets.  Comparing several kinematic distributions for
signal and background processes we devise kinematic cuts to enhance
the signal to background ratio.  The analysis reveals that the running
LHC experiment at 7 TeV (14 TeV) will explore or exclude 2 UED models
with $R^{-1}$ upto 700 GeV (1 TeV).

There are a few things which need to be mentioned here.  The signal we
have investigated in this article is, in a sense, complementary to
that in Ref. \cite{Ghosh:2008ji}, in which the production and decay of
$(1,0)$--mode electroweak particles have been discussed.
Although it is certainly true that their signal (leptons with multijets 
and missing transverse energy) has a better mass reach, it 
should be recognized that the said signal is generic to a wide 
variety of new physics scenarios. And while such models can, in principle,
be distinguished from most of the popular supersymmetric scenarios 
by virtue of the relatively closely packed spectrum, the said signal 
would also arise in the one-dimensional UED model as well.
It is here that the virtue of the hard photon in our final state 
(arising from the decay $B^{(1,0)}_\mu \rightarrow B^{(1,0)}_H + \gamma$)
lies. As such a hard photon would not arise in the minimal UED 
models, this affords us a possible means of distinguishing between
these two scenarios.

At this stage, it would not be very irrelevant to comment on a very 
similar model \cite{UED-nandi}, in which mUED is embedded in a higher
dimensional manifold with a flat geometry. As a result, $B^{(1)}_\mu$,
the lightest of the SM KK-excitations, 
is not stable any more, but decays to a photon and a graviton
with a 100\% branching ratio. Consequently, in such a model, two hard
photons emerge in the final state, with missing energy resulting
from the gravitons. 
While such a final state can occur in the 2UED scenario as well,
(if both decay chains end in $B_\mu^{(1,0)} \to B_H^{(1,0)} + \gamma$), 
note that this particular chain has a relatively small 
cumulative branching fraction. 
As a result, final states with a single photon (along with 
multijets and missing transverse energy) will far outnumber 
those with a pair of such photons. Such a hierarchy in the 
final state is quite the opposite of the situation 
of Ref.\cite{UED-nandi} , thereby allowing to differentiate between
these two scenarios as well\footnote{The situation for gauge mediated 
supersymmetry breaking models is similar to that in Ref.\cite{UED-nandi},
except that the spectrum is not as degenerate.}.

Finally, we have shown that there exists a correlation between the
peak position of $M_{eff}$ distribution and $R^{-1}$. However,
the above peak position also depends on the cut-off scale
$M_s$. Consequently, it would not be possible to determine $R^{-1}$ or
$M_s$ only by measuring the $M_{eff}$ distribution. An umambiguous
measurement of parameters thus calls for further experimental
measurement of quantities depending on these two parameters.

\section{Acknowledgments} DC acknowledges illuminating discussions with 
Mihoko Nojiri, Yasuhiro Okada and Kohsuke Tobioka. 
AD acknowledges partial financial support from 
the UGC-DRS programme of the Department of Physics, Calcutta
University. DKG acknowledges partial support from the
Department of Science and Technology, India under the grant SR/S2/HEP-12/2006.
DKG also thanks the High Energy Physics Group of ICTP for hospitality at a 
later stage of the project. KG was partially supported by funding 
available from the Department of Atomic Energy, India, for the Regional
Centre for Accelerator-based Particle Physics (RECAPP), Harish-Chandra
Research Institute. 
Authors would like to thank A. Kundu for pointing
out a typo in the legend of Fig.11.

\end{document}